\documentclass[12pt]{iopart}
\usepackage{graphicx}
\usepackage{bm}
\usepackage{color}

\def\l{\langle}
\def\r{\rangle}
\def\XY{XY}
\newcommand{\myeq}[1]{(#1)}
\newcommand{\mytable}[0]{table}
\newcommand{\mytables}[0]{tables}
\newcommand{\myfig}[0]{figure}
\newcommand{\myfigs}[0]{figures}
\newcommand{\myref}[0]{}

\begin{document}

\title[Comprehensive studies on BKT transitions]{Comprehensive studies on the universality of BKT transitions --- Machine-learning study, Monte Carlo simulation, and Level-spectroscopy method}

\author{Hiromi Otsuka$^1$, Kenta Shiina$^{2,1}$, and Yutaka Okabe$^1$}

\address{$^1$Department of Physics, Tokyo Metropolitan University, Hachioji, Tokyo 192-0397, Japan\\$^2$AiCAN Inc., Kanagawa Science Park, 3-2-1 Sakado, Takatsu-ku, Kawasaki 213-0012, Japan}
\ead{otsuka@tmu.ac.jp}
\vspace{10pt}
\begin{indented}
  \item[]\today
\end{indented}
\begin{abstract}
  Comprehensive studies are made on the six-state clock universality
  of two models using several approaches.
  We apply the machine-learning technique of phase classification
  to the antiferromagnetic (AF) three-state Potts model
  on the square lattice with ferromagnetic next-nearest-neighbor (NNN)
  coupling and the triangular AF Ising model with anisotropic NNN coupling
  to study two Berezinskii-Kosterlitz-Thouless transitions.
  We also use the Monte Carlo simulation paying attention to the ratio of
  correlation functions of different distances for these two models.
  The obtained results are compared with those of the previous
  studies using the level-spectroscopy method.
  We directly show the six-state clock universality
  for totally different systems with the machine-learning study.
\end{abstract}
\vspace{10pt}
\begin{indented}
  \item[]Keywords: BKT transitions, Machine-learning, Monte Carlo, Level-spectroscopy
\end{indented}

\maketitle

\section{Introduction}

The concepts of scaling and universality have played an essential role
in statistical physics, especially in the study of
phase transitions~\cite{Kadanoff,Kadanoff1990}.
There are sometimes implicit symmetries in the models of physics.
Universality appears in totally different systems.
One example is a six-state clock universality.

The two-dimensional (2D) spin systems with a continuous {\XY} symmetry
exhibit a unique phase transition called the
Berezinskii-Kosterlitz-Thouless (BKT) transition
\cite{Berezinskii1,Berezinskii2,kosterlitz,kosterlitz2}.
A BKT phase of a quasi-long-range order, a fixed line,
exists, wherein the correlation function decays as a power law.
The $q$-state clock model, which is a discrete version
of the {\XY} model, experiences two BKT transitions
for $q>4$ due to the discreteness~\cite{jose,elitzur}.
It is noteworthy that in the clock model with a modified interaction
of Villain type~\cite{villain}, there is an exact
duality relation between the two (higher and lower temperatures)
transitions.

Antiferromagnetic (AF) spin systems may have a variety of
phase transitions compared to ferromagnetic (F) counterparts.
The F three-state Potts model on the square lattice exhibits
a second-order phase transition with $T_c$ of $1/\ln(1+\sqrt{3})$
in units of the coupling $J$.
On the contrary, the AF three-state Potts model on the
square lattice does not possess the order, because there are
many possibilities to choose different states for
the nearest-neighbor (NN) pairs.
%\textcolor{red}{
We note that these features are not only present 
in the three-state AF Potts model on the square lattice but are 
also present in the three-state AF Potts model 
on a large class of plane quadrangulations~\cite{Lv}.
%}
There is a macroscopic degeneracy in the ground state.  However,
if the F next-nearest-neighbor (NNN) coupling is added~\cite{den_Nijs},
diagonal spins take the same state.
That is, all the spins in the sublattice "a" take the same state,
one of three, and all the spins in the sublattice "b" take one of
the other two states.  Then, the degeneracy of the ground state
becomes six ($3 \times 2$). This model is thought to be
in the same universality class as the F six-state clock
model in the analysis of essentially the same model
with three-fold symmetry-breaking fields~\cite{cardy1981}.
Another example is the AF Ising model on the triangular lattice
with an anisotropic NNN coupling, which was introduced by
Kitatani and Oguchi~\cite{kitatani}.
A detailed description of the anisotropic NNN coupling
will be given in Sec. II.  There is no order in the
AF Ising model on the triangular lattice without NNN couplings
due to frustration, and a macroscopic degeneracy appears
in the ground state. With the introduction of the anisotropic NNN
coupling, spins are classified into interpenetrating three sublattices,
and the system is considered to belong to the six-state clock
universality class~\cite{kitatani}.

There is a difficulty in the numerical analysis of BKT transitions.
The divergence of the correlation length for the BKT transition is more rapid than any power law, and there are logarithmic corrections.
These pathological features make it difficult to determine the BKT transition point from finite-size calculations.
Nomura proposed a method to determine BKT points observed in quantum spin chains with high precision~\cite{nomuraOkamoto}, the so-called level spectroscopy (LS) method~\cite{nomura}.
It was pointed out that an enhancement of the U(1) symmetry of the 2D Gaussian critical model to an SU(2) symmetry of a Wess-Zumino-Witten model brings about a typical degeneracy in an excitation spectrum, which then can be detected as a crossing of excitation levels of constituent physical quantities to an SU(2) multiplet~\cite{nomuraOkamoto}.
Although the levels to analyze depend on the types of BKT transitions, we can specify them in numerical calculations of finite-size systems with the help of the symmetry properties of physical quantities and thus can determine BKT transition points.
Nakamura~{\it et al.} extended the LS method to apply to the one-dimensional electron systems; for example, see \myref\cite{nakamura}.
Alternatively, Otsuka~{\it et al.} studied a 2D classical spin model with frustration, the AF three-state Potts model on the square lattice with F NNN coupling~\cite{Otsuka}.
Combining the exact diagonalization calculations and the LS analysis, they showed two BKT transitions and the universality with the F six-state clock model (see Sec.~II).
The precise phase diagram in the space of the NN coupling and
the NNN coupling was obtained.
The triangular AF Ising model with anisotropic NNN coupling
was also studied by Otsuka~{\it et al.}~\cite{Otsuka2}
by the LS method.
The six-state clock universality was discussed.
Although some numerical works on these two models had been
reported previously, the LS studies~\cite{Otsuka,Otsuka2}
gave reliable global phase diagrams.

The Monte Carlo (MC) simulation has served as a standard method
for the numerical analysis of many particle systems~\cite{Landau}.
The finite-size scaling (FSS) study~\cite{Barber,Cardy} of
the Binder ratio~\cite{Binder}, essentially the moment ratio, is a powerful tool for the
second-order phase transition.
However, for the BKT transition, the Binder ratio is not
so effective~\cite{hasenbusch} because of the multiplicative
logarithmic corrections~\cite{kosterlitz2,Janke}.
There are other quantities that have the same FSS behavior
with a single variable as the Binder ratio.
The ratio of the correlation functions of different
distances~\cite{Tomita2002} is one example.
The size-dependent second-moment correlation length
for a finite system, $\xi(L)$, has been frequently used
in spin glass problems~\cite{katzgraber}.
The FSS of $\xi(L)/L$ also has the same form of a single
scaling variable.
Surungan {\it et al.}~\cite{Surungan} made the MC study
of the correlation ratio and the size-dependent second-moment
correlation length for the $q$-state clock model
of both the standard cosine-type interaction and the Villain-type
interaction.  They observed the collapsing curves
of different sizes at intermediate temperature regime
and the spray out at lower and higher temperatures,
which demonstrates two BKT transitions.

In addition to conventional approaches, new numerical strategies
are developing.
Recent advances in machine-learning-based techniques
have been applied to fundamental research, such as
statistical physics~\cite{Carleo}.
Carrasquilla and Melko~\cite{Carrasquilla} used a technique
of supervised learning for image classification,
which is complementary to the conventional approach
of studying interacting spin systems.
They classified and identified a high-temperature paramagnetic phase
and a low-temperature ferromagnetic phase of
the 2D Ising model by using data sets of spin configurations.
Shiina~{\it et al.}~\cite{Shiina} extended and generalized
this machine-learning approach to studying various spin models
including the multi-component systems and the systems
with a vector order parameter.
The configuration of a long-range spatial correlation
was treated instead of the spin configuration itself.
Not only the second-order and the first-order transitions
but also the BKT transition was studied.
They detected two BKT transitions for the six-state clock model.
Tomita {\it et al.}~\cite{Tomita2020} made further progress
in the machine-learning study of phase classification.
%\textcolor{red}{
When the cluster update is possible in the MC simulation,
the Fortuin-Kasteleyn (FK) \cite{KF,FK}
representation-based improved estimators
\cite{Wolff90,Evertz93} for the configuration of two-spin correlation
were employed as an alternative to the ordinary spin configuration.
%} 
This method of improved estimators was applied not only
to the classical spin models but also to the quantum MC simulation.
Miyajima~{\it et al.} also used a machine-learning approach to study
two BKT transitions of the eight-state clock model~\cite{Miyajima}.
%\textcolor{red}{
We note that the principal component analysis was used in 
the machine-learning study on the {\XY} model~\cite{Hu,Wang}.
%}
As another approach to machine learning, the image super-resolution
using the deep convolutional neural networks was applied to the study of
the inverse renormalization group of spin systems~\cite{Shiina2021}.

Quite recently, much attention has been paid to the study of
the $q$-state clock models~\cite{Li,Hong,Ueda,Li2,Chen}.
Some works used newly developed methods, such as the tensor renormalization
methods~\cite{Li,Hong,Li2}, the corner-transfer matrix renormalization
group~\cite{Ueda}, and the worm-type simulation~\cite{Chen}.

In the present paper, we perform the comprehensive studies
on two spin models, the AF three-state Potts model
on the square lattice with F NNN coupling and
the triangular AF Ising model with anisotropic NNN coupling.
We first apply the machine-learning approach to these models.
Using the training data of the F six-state clock model,
we demonstrate the classification of the disordered,
the BKT, and the ordered phase.
We then perform the MC studies on two models.
Analyzing the correlation ratios, we examine two BKT transitions.
Finally, we compare the results with the LS method
by Otsuka~{\it et al.}~\cite{Otsuka,Otsuka2}.

The rest of the paper is organized as follows:
In Sec.~II, we explain the models and simulation models.
Sections~III and IV are devoted to the results of
the AF three-state Potts model
on the square lattice with F NNN coupling and
the triangular AF Ising model with anisotropic NNN coupling,
respectively.  The summary and discussion are given
in Sec.~V.  The calculations of the {\XY} and six-state clock models
on the triangular lattice are presented in the Appendix.

\section{Models and numerical methods}

We treat two spin models in the present study. The first one is
the AF three-state Potts model on the square lattice
with F NNN coupling~\cite{den_Nijs},
whose Hamiltonian is given by
\begin{equation}
  H = J_1 \sum_{\l ij \r} \delta_{s_i s_j}
  - J_2 \sum_{[ ij ]} \delta_{s_i s_j}, \quad s_i = 1, 2, 3,
  \label{Potts}
\end{equation}
where $\delta_{a b}$ is the Kronecker delta.
The first and the second sums run over all NN
and NNN pairs, respectively.
The periodic boundary conditions are imposed in numerical
simulations.
The couplings $J_1$ and $J_2$ are positive, and hereafter
we use $r = J_2/J_1$.  The temperature will be measured
in units of $J_1$.

\begin{figure}
  \begin{center}
    \includegraphics[width=7cm]{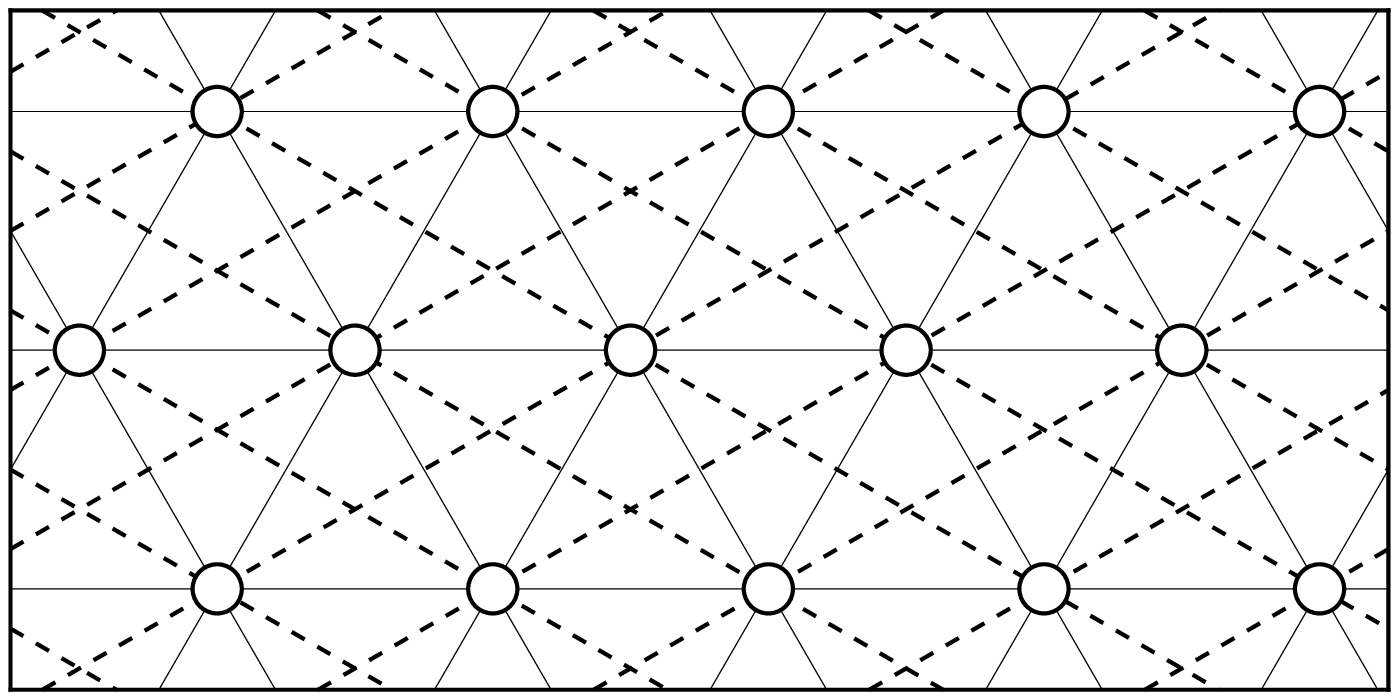}
    \caption{
      The schematic representation of Hamiltonian~(\ref{Tri_Ising})
      on the triangular lattice~\cite{Otsuka2}.
      The NN couplings are denoted by straight lines, whereas
      the anisotropic NNN couplings in two of the three directions
      are denoted by dotted lines.
    }
    \label{fig:NNNinteraction}
  \end{center}
\end{figure}

The second model is the AF Ising model on the triangular lattice
with anisotropic NNN coupling, which was studied by Kitatani and
Oguchi~\cite{kitatani}. The Hamiltonian is given by
\begin{equation}
  H =
  J_1 \sum_{\l ij \r} \delta_{s_i s_j}
  - J_2 \sum_{[ ij ]'} \delta_{s_i s_j}, \quad s_i = 1, 2,
  \label{Tri_Ising}
\end{equation}
where the first sum runs over all NN pairs,
whereas the second sum runs over anisotropic NNN pairs
in two of the three directions.
The schematic representation of Hamiltonian is given
in \myfig~\ref{fig:NNNinteraction}.
The periodic boundary conditions are imposed in numerical
simulations.
The couplings $J_1$ and $J_2$ are positive, and
we use $r = J_2/J_1$.  The temperature will be measured
in units of $J_1$ again.

%%%%%%%%%%%%%%%%%%%%%%%%%%%%%%%%%%%%%%%%%%%%%%%%%%%%%%%%%%%%%%%%%%%%%%%%%%%%
%%%%%%%%%%%%%%%%%%%%%%%%%%%%%%%%%%%%%%%%%%%%%%%%%%%%%%%%%%%%%%%%%%%%%%%%%%%%

The correlation function with the distance $r$
for the $q$-state Potts model including the Ising model
($q=2$) is given by
\begin{equation}
  g_i(r) = \frac{q \delta_{s_i s_{i+r}} - 1}{q-1}.
\end{equation}
It assumes a value of $+1$ or $-1/(q-1)$.

We use three approaches to study the BKT transitions.  The first approach
is the machine-learning study which was developed by
Shiina~{\it et al.}~\cite{Shiina},
to classify the ordered, the BKT, and the disordered phases.
A fully connected neural network is implemented with a standard library
of TensorFlow of the 100-hidden unit model.
For the spin data, we use the long-range correlation configuration
with the distance of $L/2$ instead of the spin configuration itself.
The average of the correlation along the $x$-axis and the $y$-axis
is considered.  For the training data, we use the results of
the F six-state clock model~\cite{Shiina}.  For the test data,
we investigate the correlation configurations of the AF three-state
Potts model with F NNN coupling and that of the triangular AF Ising model with anisotropic NNN coupling.
We classify the phases of these two models
using the training data of the six-state clock model.

As the second approach, we perform the MC simulation
of the ratio of the correlation functions of different distances,
$R(T)=\l g(L/2) \r/\l g(L/4) \r$; for the two distances,
we chose $L/2$ and $L/4$.
%\textcolor{red}{
Surungan~{\it et al.}~\cite{Surungan} made MC studies
of the BKT transitions by investigating the correlation ratio
and the size-dependent second-moment correlation length. 
Although there are other candidate quantities, 
we take the correlation ratio here.
%}
The correlation ratio has a single scaling variable
for FSS
\begin{equation}
  R(T) = \frac{\l g(L/2) \r}{\l g(L/4) \r} = \tilde f(L/\xi),
\end{equation}
as in the Binder ratio~\cite{Binder}, which is essentially the ratio of
different moments. Here, $\xi$ stands for the correlation length.
At the critical region, where the correlation length $\xi$ is infinite,
the correlation ratio does not depend on the system size $L$.
Thus, we expect the data collapsing of different sizes in the BKT phase.
Above the upper BKT temperature $T_2$ and below the lower
BKT temperature $T_1$, the data of different sizes start to separate.
In the case of the BKT transition, the correlation length
diverges as
\begin{equation}
  \xi \propto \exp (c/\sqrt{|t|}),
\end{equation}
where $t = T-T_{1,2}$.
%\textcolor{red}{
We employed the Metropolis local-update algorithm for the two models 
in combination with the replica-exchange algorithm~\cite{Hukushima} 
that overcomes slow dynamics. For the clock model we used 
the Swendsen-Wang cluster-update algorithm~\cite{SW} in combination 
with the replica-exchange algorithm. 
The typical number of Monte Carlo steps was 200000 after discarding 
the first 10000 steps. Since the temperature exchange was typically performed 
for 40 temperatures, the effective Monte Carlo step number is much larger.
%}

The third approach is the LS method employed in the previous publications~\cite{Otsuka,Otsuka2}.
To make the explanations concrete, we shall borrow some notations there.
Based on a dual sine-Gordon model, an effective field theory of the lattice models
(see \myeq{2} in \myref\cite{Otsuka} and \myeq{4} in \myref\cite{Otsuka2}), we analyzed the scaling dimensions of some local operators.
For the upper BKT transition, the so-called ${\cal M}$ operator (a part of the Gaussian term) hybridizes with an operator $\sqrt2\cos\!\sqrt2\theta$ given in the disorder field and yields the ${\cal M}$-like operator with the dimension ${\bar x}_0$ \cite{nomura}.
We also focused on an operator $\sqrt2\sin3\sqrt2\phi$ given in the order field with the dimension $x_2$.
The numerical analysis of the eigenvalues of transfer matrices supposes a lattice model with a $L\times\infty$ cylindrical  geometry.
As explained in Refs.~\cite{Otsuka,Otsuka2}, we can evaluate the renormalized scaling dimensions ${\bar x}_0(l)$ and $x_2(l)$ ($l=\ln L$) using discrete symmetry properties of these operators.
We then obtained finite-size estimates of $T_2$ by the level-crossing condition
\begin{equation}
  {\bar x}_0(l)=\frac{16}9x_2(l).
\end{equation}
For the lower BKT transition, the ${\cal M}$ operator hybridizes with $\sqrt2\cos6\sqrt2\phi$ and yields another ${\cal M}$-like operator with $x_0$
\cite{nomura}.
We thus calculated finite-size estimates of $T_1$ by the level-crossing condition
\begin{equation}
  x_0(l)=4x_2(l).
\end{equation}

We compare the estimated BKT temperatures $T_{1,2}$ by three approaches.

\section{AF three-state Potts model with ferromagnetic NNN coupling}

\subsection*{Machine-learning study}

\begin{figure}
  \begin{center}
    \includegraphics[width=5.0cm]{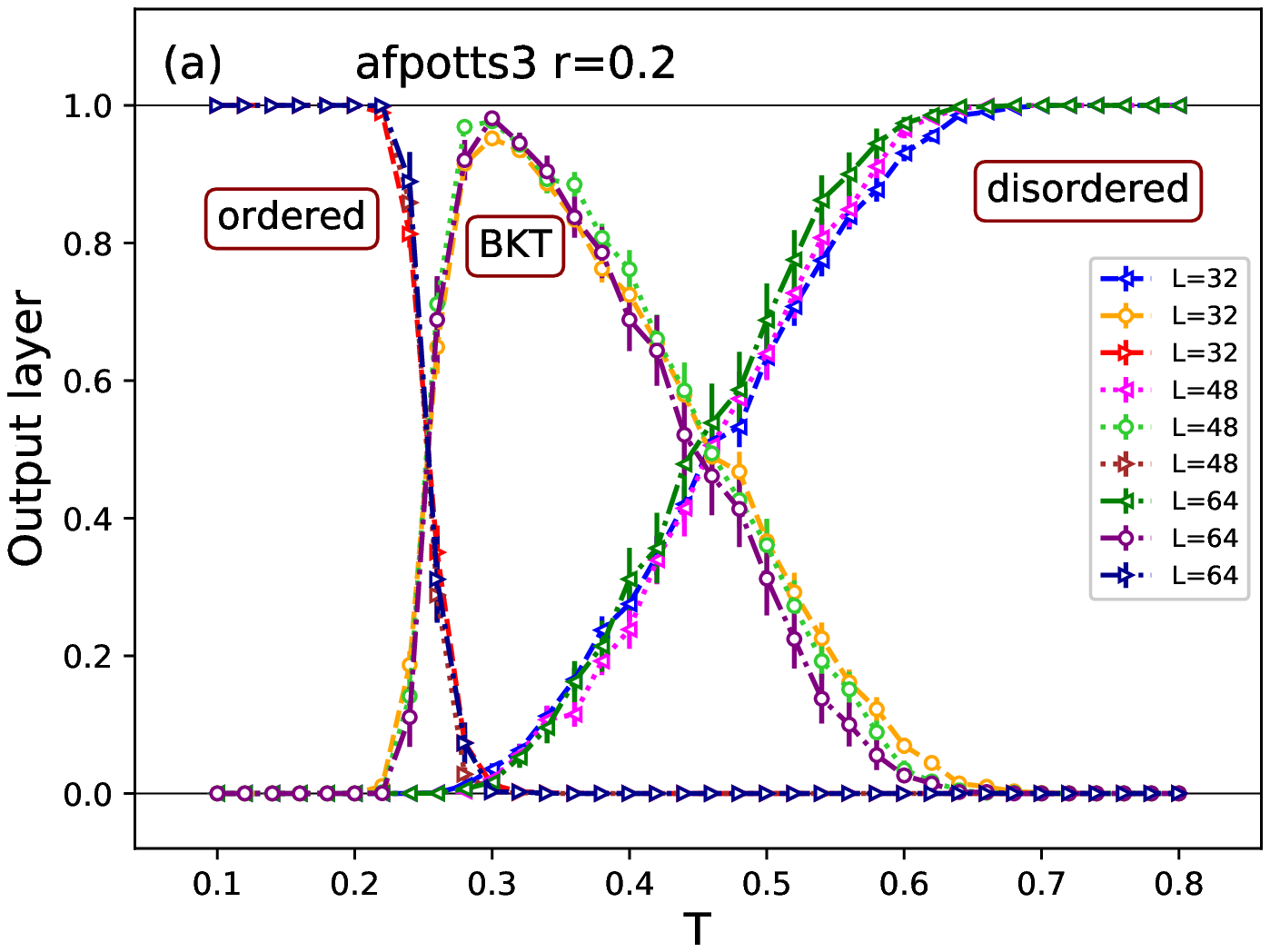}\hfill
    \includegraphics[width=5.0cm]{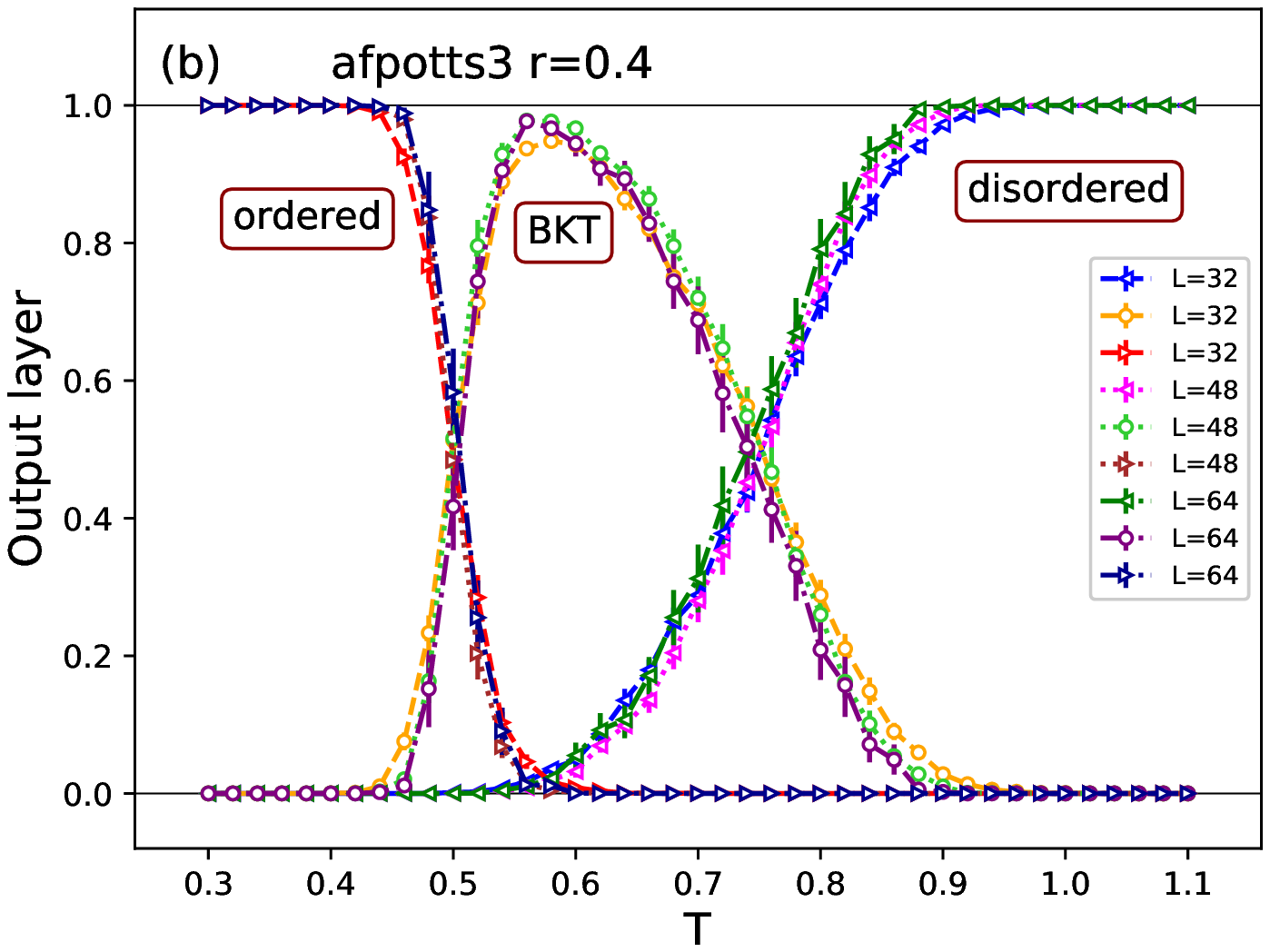}\hfill
    \includegraphics[width=5.0cm]{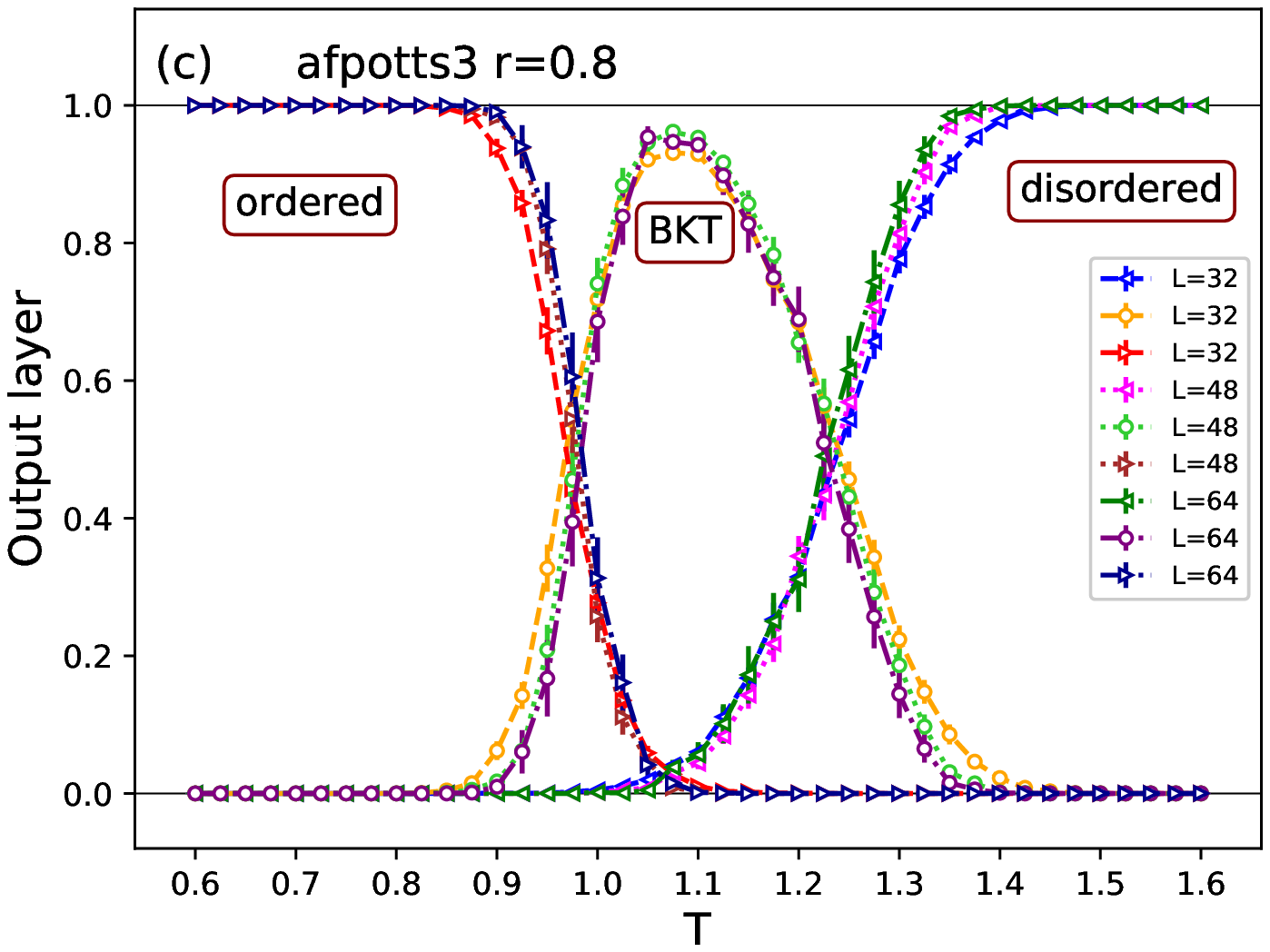}
    \caption{
      The output layer for the AF three-state Potts model with F NNN coupling
      for $r=0.2$ (a), $r=0.4$ (b), and $r=0.8$ (c),
      using the training data of the F six-state clock model.
      The system sizes are $L = 32$, 48, and 64.
    }
    \label{fig:output_AFpotts3}
  \end{center}
\end{figure}

\begin{figure}
  \begin{center}
    \includegraphics[width=5.0cm]{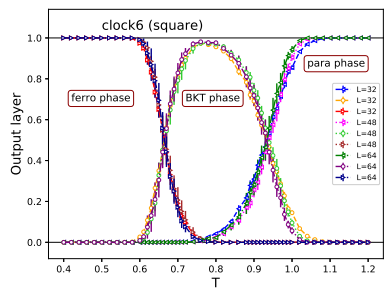}
    \caption{
      The output layer averaged over a test set as a function of $T$
      for the 2D six-state clock model on the square lattice.
      The system sizes are $L$ = 32, 48, and 64.
      The samples of $T$ within the ranges
      $0.4 \le T \le 0.64$, $0.78 \le T \le 0.82$ and
      $0.96 \le T \le 1.2$ are used for training data.
    }
    \label{fig:output_clock6}
  \end{center}
\end{figure}

We first show the results of the AF three-state Potts model
with F NNN coupling. As a machine-learning study,
the output layer averaged over a test set as a function of $T$
is plotted in \myfig~\ref{fig:output_AFpotts3}.
The data for $r$ = 0.2, 0.4, and 0.8 are given in (a), (b), and (c),
respectively. The system sizes are 32, 48, and 64.
For the training data, we use the F six-state clock model.
We show the result of the previous study~\cite{Shiina}
for the F six-state clock model in \myfig~\ref{fig:output_clock6},
for convenience. The probabilities of predicting the phases
are plotted at each temperature.
The samples of $T$ within the ranges $0.4 \le T \le 0.64$,
$0.78 \le T \le 0.82$ and
$0.96 \le T \le 1.2$ are used for the low-temperature,
middle-temperature, and high-temperature training data, respectively.
For a whole temperature range, around 35000 training data sets
are used, and we use 500 test data sets for each temperature.

%%%%%%%%%%%%%%%%%%%%%%%%%%%%%%%%%%%%%%%%%%%%%%%%%%%%%%%%%%%%%%%%%%%%%%%%%%%%
%%%%%%%%%%%%%%%%%%%%%%%%%%%%%%%%%%%%%%%%%%%%%%%%%%%%%%%%%%%%%%%%%%%%%%%%%%%%

%%%%%%%%%%%%%%%%%%%%%%%%%%%%%%%%%%%%%%%%%%%%%%%%%%%%%%%%%%%%%%%%%%%%%%%%%%%%
%%%%%%%%%%%%%%%%%%%%%%%%%%%%%%%%%%%%%%%%%%%%%%%%%%%%%%%%%%%%%%%%%%%%%%%%%%%%

Figure \ref{fig:output_AFpotts3} clearly shows the existence
of three phases, the ordered phase, the intermediate phase, and
the disordered phase.
We estimate the size-dependent $T_{1,2}(L)$ from the point
that the probabilities of predicting two phases are 50\%.
Systematic size dependence is not observed.
The rough estimates of the transition temperatures,
$T_1$ and $T_2$, are tabulated in \mytable~\ref{table_AFPotts}.
We emphasize that the BKT transitions of a complex model,
AF three-state Potts model on the square lattice with F NNN
coupling can be studied by the training data of the
six-state clock model.

%%%%%%%%%%%%%%%%%%%%%%%%%%%%%%%%%%%%%%%%%%%%%%%%%%%%%%%%%%%%%%%%%%%%%%%%%%%%
\begin{figure}
  \begin{center}
    \includegraphics[width=5.0cm]{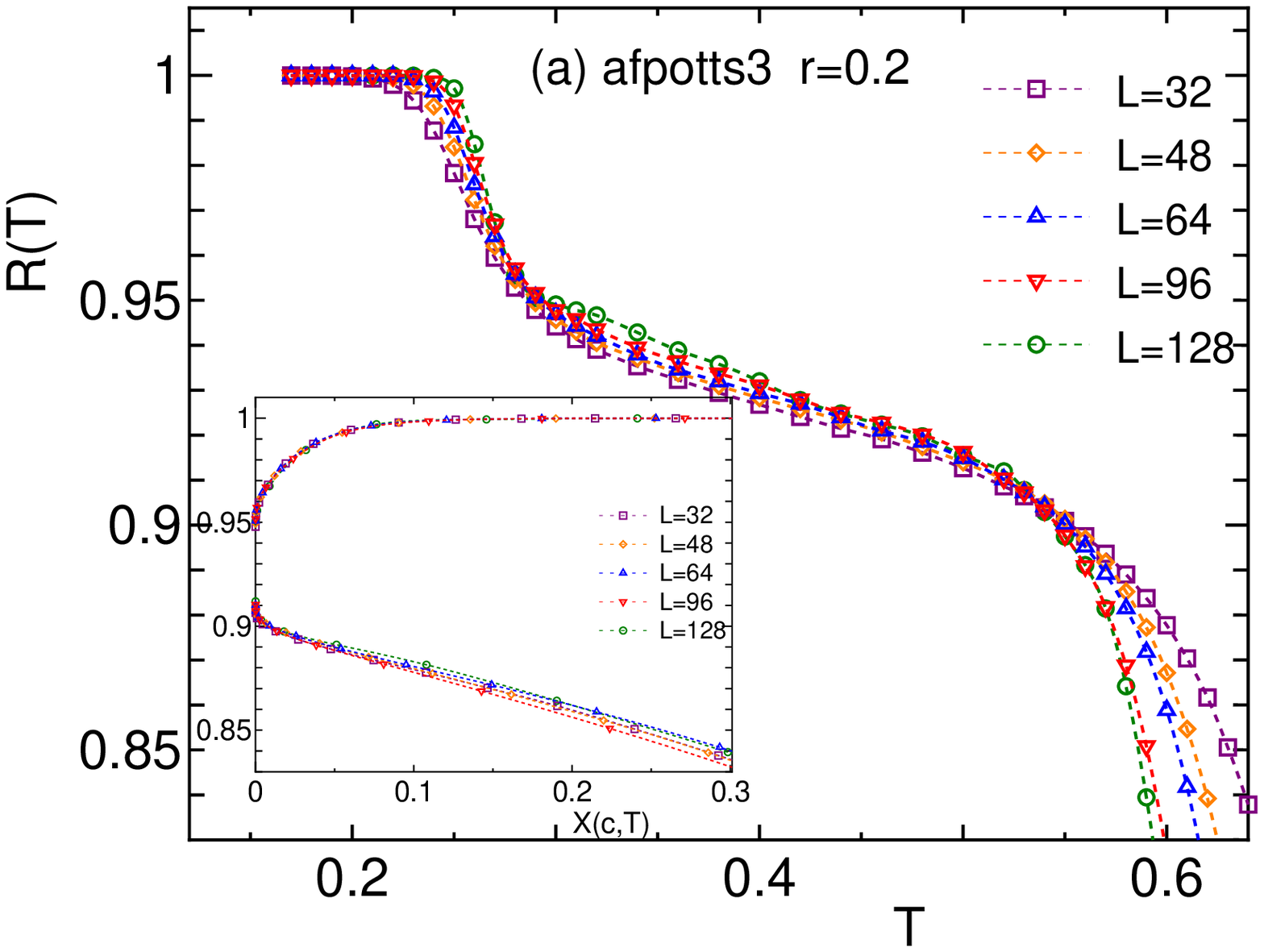}\hfill
    \includegraphics[width=5.0cm]{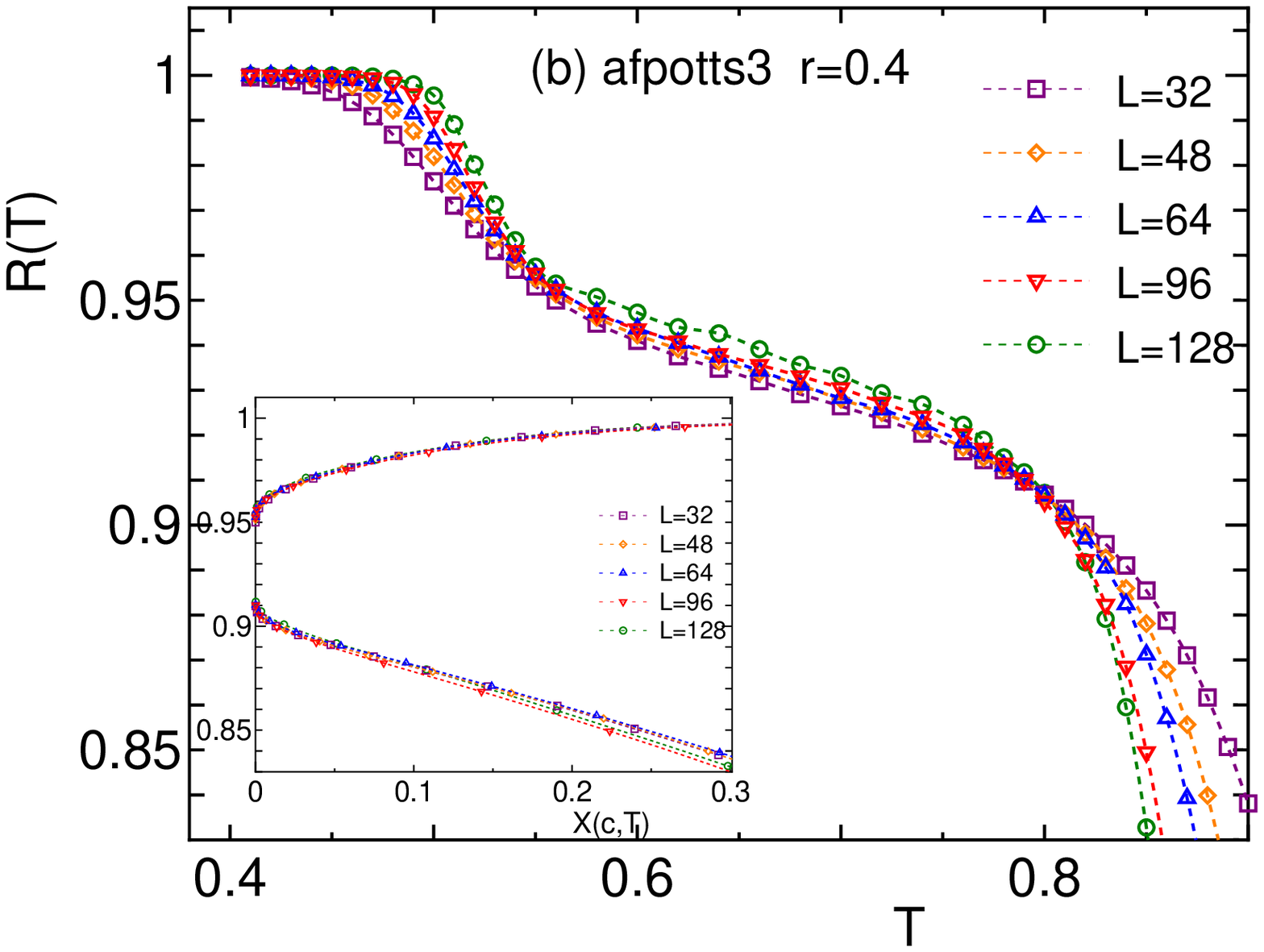}\hfill
    \includegraphics[width=5.0cm]{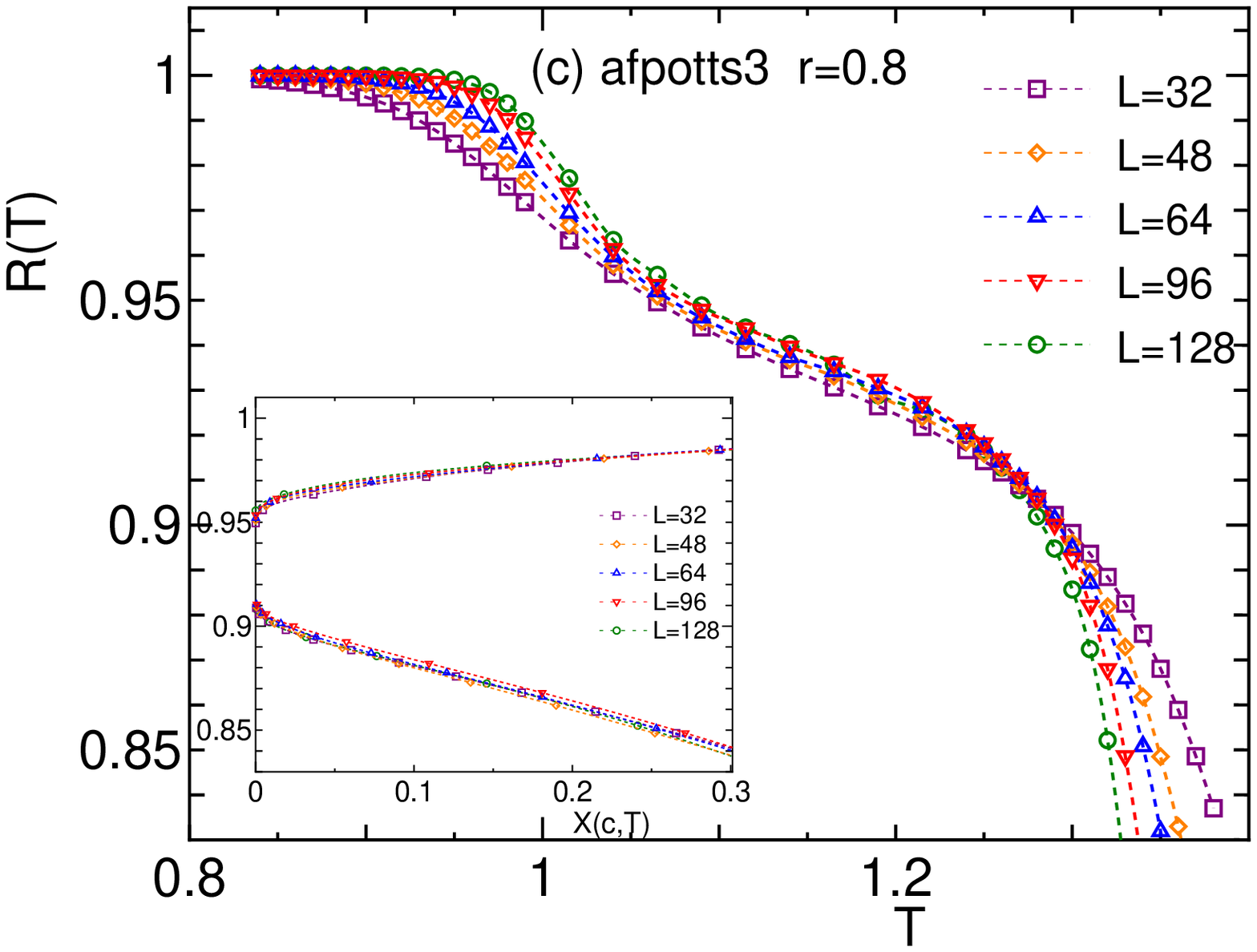}
    \caption{
      The plot of the correlation ratio $R(T)$ for the AF three-state Potts model
      on the square lattice with F NNN coupling.
      In the inset, the FSS plots are given,
      where $X(c,T)=L/\exp(c_{1,2}/\sqrt{|T-T_{1,2}|})$.
    }
    \label{fig:MC_AFPotts}
  \end{center}
\end{figure}
%%%%%%%%%%%%%%%%%%%%%%%%%%%%%%%%%%%%%%%%%%%%%%%%%%%%%%%%%%%%%%%%%%%%%%%%%%%%

\subsection*{Monte Carlo study}

We here show the MC results of the AF three-state Potts model
on the square lattice with F NNN coupling.
Following the study by Surungan {\it et al.}~\cite{Surungan},
we calculate the correlation ratio, $R(T)=\l g(L/2) \r/\l g(L/4) \r$,
to investigate the BKT transitions.
The results for $r$ = 0.2, 0.4, and 0.8 are plotted in
\myfig~\ref{fig:MC_AFPotts}.  The system sizes are $L=32$, 48,
64, 96, and 128. We observe the collapsing curves of different sizes
at intermediate temperature regimes
and the spray out at lower and higher temperatures,
which indicates the existence of two BKT transitions.
The behavior of the collapsing is not good enough,
compared to the regular six-state clock model~\cite{Surungan},
which may be attributed to the large corrections to FSS.

In the case of the BKT transition, the correlation length
diverges as
\begin{equation}
  \xi \propto \exp (c/\sqrt{|t|}),
\end{equation}
where $t = T-T_{1,2}$.
Thus, we can get the FSS plot if the temperature $T$ is scaled as $L/\exp(c_{1,2}/\sqrt{|T-T_{1,2}|})$, where $c$ is a fitting parameter.
The FSS plots are given in the inset of \myfig~\ref{fig:MC_AFPotts}.

The rough estimates of $T_1$ and $T_2$
are also tabulated in \mytable~\ref{table_AFPotts}.

\subsection*{Level-spectroscopy study}

\begin{figure}
  \begin{center}
    \includegraphics[width=7cm]{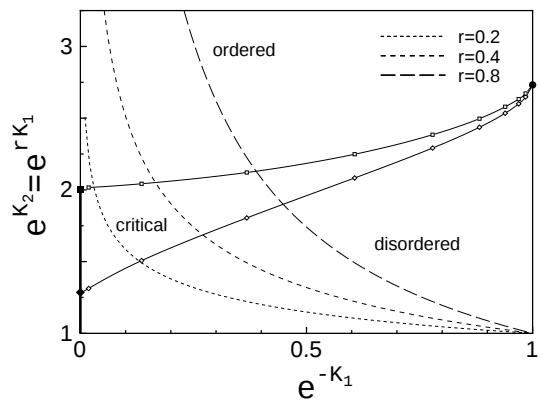}
    \caption{
    The phase diagram of the AF three-state Potts model
    on the square lattice with F NNN coupling
    using the LS method~\cite{Otsuka}
    in the space of $(u,v)$ = $(e^{-K_1}, e^{K_2})$.
    The curves $v=e^{-r\ln (u)}$ with $r = 0.2$, 0.4, and 0.8
    are also plotted.
    From the crossing point with these curves,
    we can obtain the estimates of $T_1$ and $T_2$.
    }
    \label{fig:otsuka_AFPotts}
  \end{center}
\end{figure}

This model was studied using the LS method~\cite{Otsuka}.
The phase diagram, in the space of $u=e^{-K_1} = e^{-J_1/T}$ and
$v=e^{K_2} = e^{r J_1/T}$,
was given in \myfig~2 of \myref\cite{Otsuka}.
The phase diagram is reproduced in \myfig~\ref{fig:otsuka_AFPotts}, and
the curves $v=e^{-r\ln (u)}$ with $r = 0.2$, 0.4, and 0.8 are added.
From the crossing point with these curves,
we can read off the estimates of $T_1$ and $T_2$.
The estimates are tabulated in \mytable~\ref{table_AFPotts}.

%%%%%%%%%%%%%%%%%%%%%%%%%%%%%%%%%%%%%%%%%%%%%%%%%%%%%%%%%%%%%%%%%%%%%%%%%%%%
%%%%%%%%%%%%%%%%%%%%%%%%%%%%%%%%%%%%%%%%%%%%%%%%%%%%%%%%%%%%%%%%%%%%%%%%%%%%

%%%%%%%%%%%%%%%%%%%%%%%%%%%%%%%%%%%%%%%%%%%%%%%%%%%%%%%%%%%%%%%%%%%%%%%%%%%%
\begin{table}
  \caption{
    The estimates of $T_1$ and $T_2$ for the AF three-state Potts model
    on the square lattice with F NNN coupling. The estimates of
    the LS method are based on the calculation in \myref\cite{Otsuka}.
  }
  \label{table_AFPotts}
  \begin{center}
    \begin{tabular}{lllllll}
      \hline
      \hline
                         & \multicolumn{2}{l}{$r=0.2$} 
                         & \multicolumn{2}{l}{$r=0.4$}
                         & \multicolumn{2}{l}{$r=0.8$}                                                                       \\
                         & \ $T_1$                     & \ $T_2$                     & \ $T_1$ & \ $T_2$ & \ $T_1$ & \ $T_2$ \\
      \hline
      machine-learning   & \ 0.25                      & \ 0.44                      & \ 0.50  & \ 0.74  & \ 0.98  & \ 1.23  \\
      Monte Carlo        & \ 0.30                      & \ 0.51                      & \ 0.57  & \ 0.775 & \ 1.075 & \ 1.255 \\
      level-spectroscopy & \ 0.277                     & \ 0.490                     & \ 0.554 & \ 0.755 & \ 1.056 & \ 1.240 \\
      \hline
      \hline
    \end{tabular}
  \end{center}
\end{table}
%%%%%%%%%%%%%%%%%%%%%%%%%%%%%%%%%%%%%%%%%%%%%%%%%%%%%%%%%%%%%%%%%%%%%%%%%%%%

By comparing various estimates of $T_1$ and $T_2$ tabulated
in \mytable~\ref{table_AFPotts}, we see the consistency of various methods.
The machine-learning study using the training data of the six-state clock
model yields reasonable estimates.

\section{Triangular AF Ising model with anisotropic NNN coupling}

\subsection*{Machine-learning study}

Next, we consider the triangular AF Ising model with anisotropic NNN coupling.
Because the $q$-state clock models ({\XY} model in the limit
of $q \to \infty$) on the triangular lattice
were not studied previously in the literature,
the detailed calculation of the clock models
on the triangular lattice is given in the Appendix.

\begin{figure}
  \begin{center}
    \includegraphics[width=5.0cm]{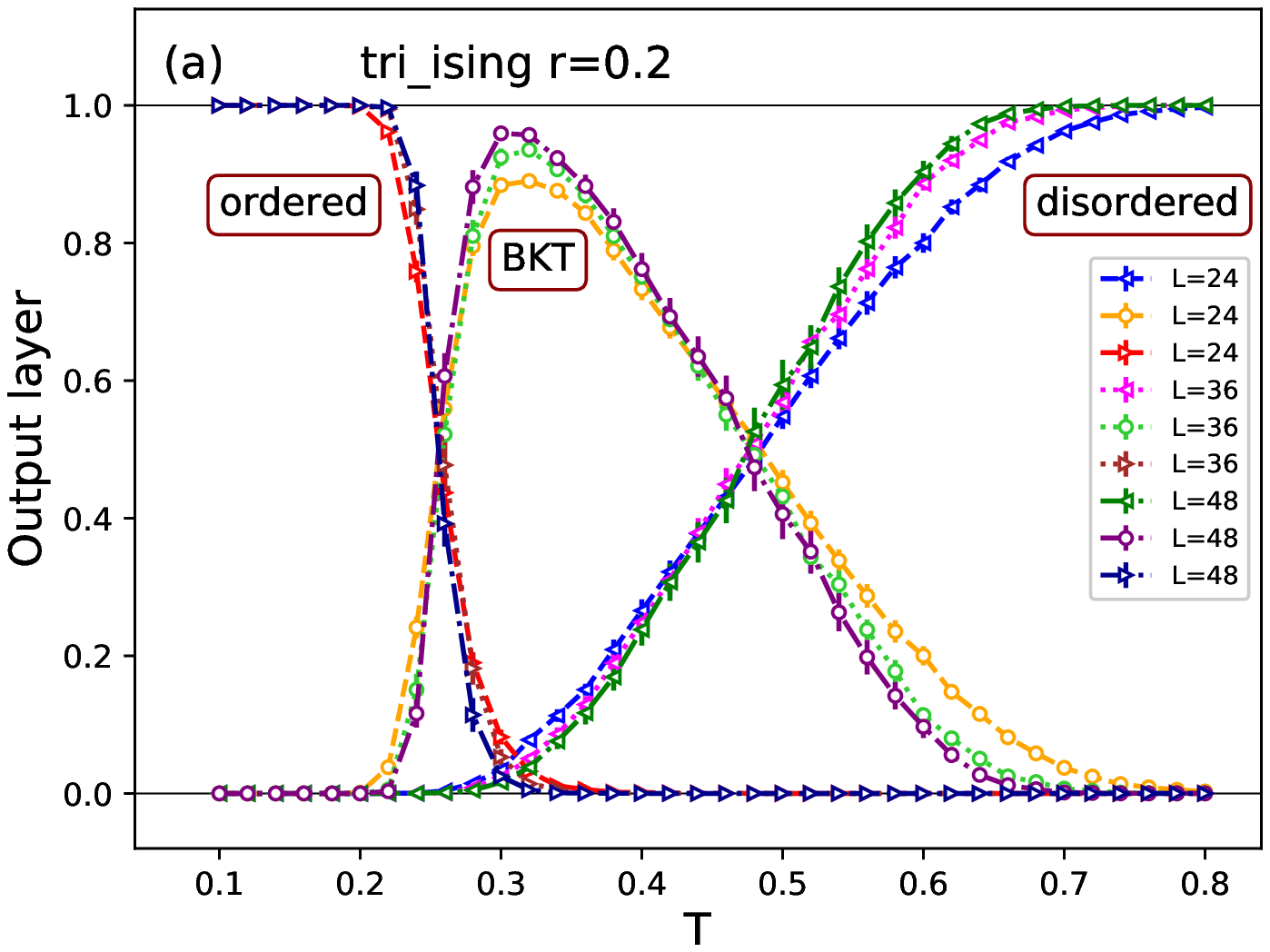}\hfill
    \includegraphics[width=5.0cm]{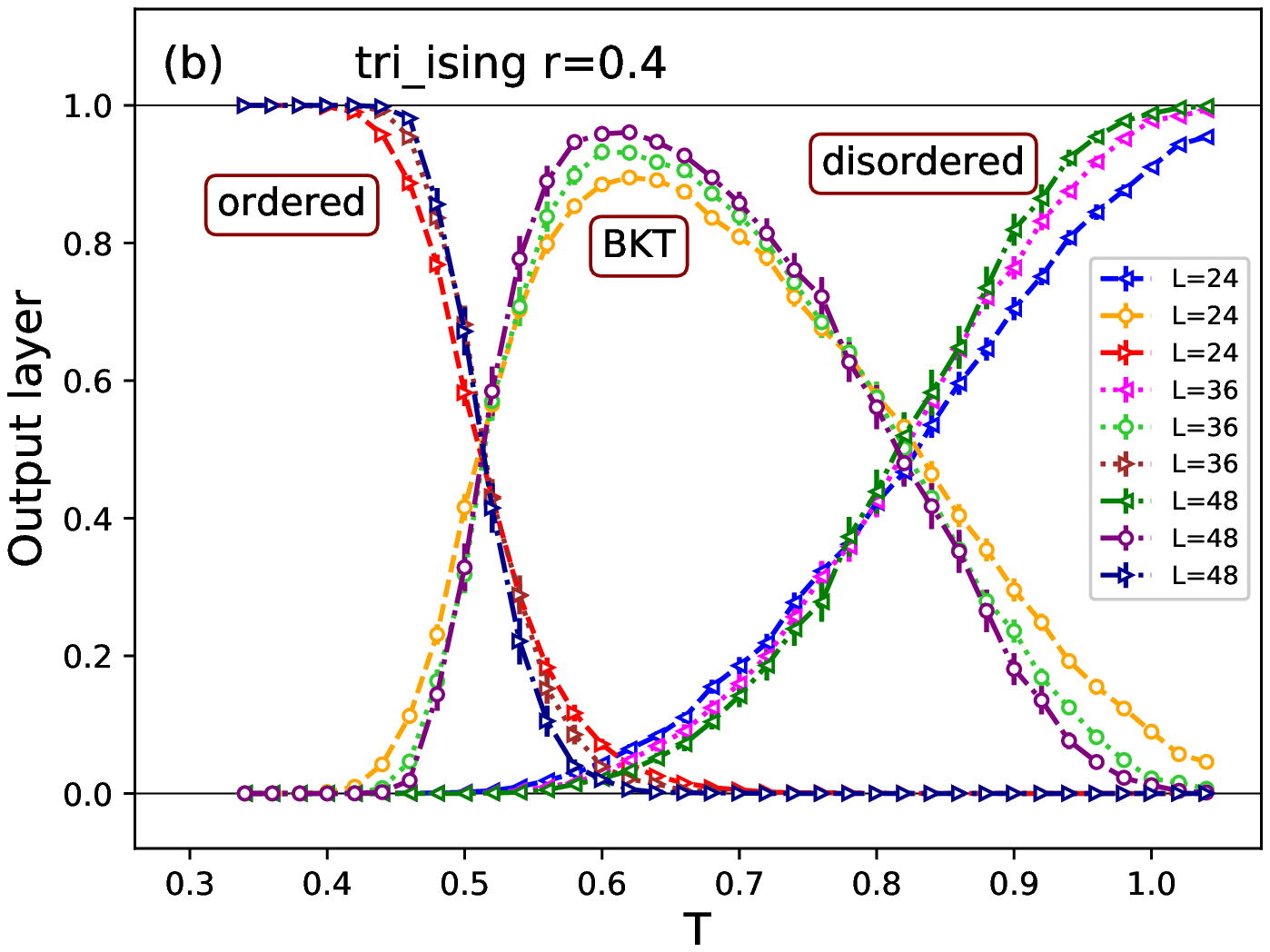}\hfill
    \includegraphics[width=5.0cm]{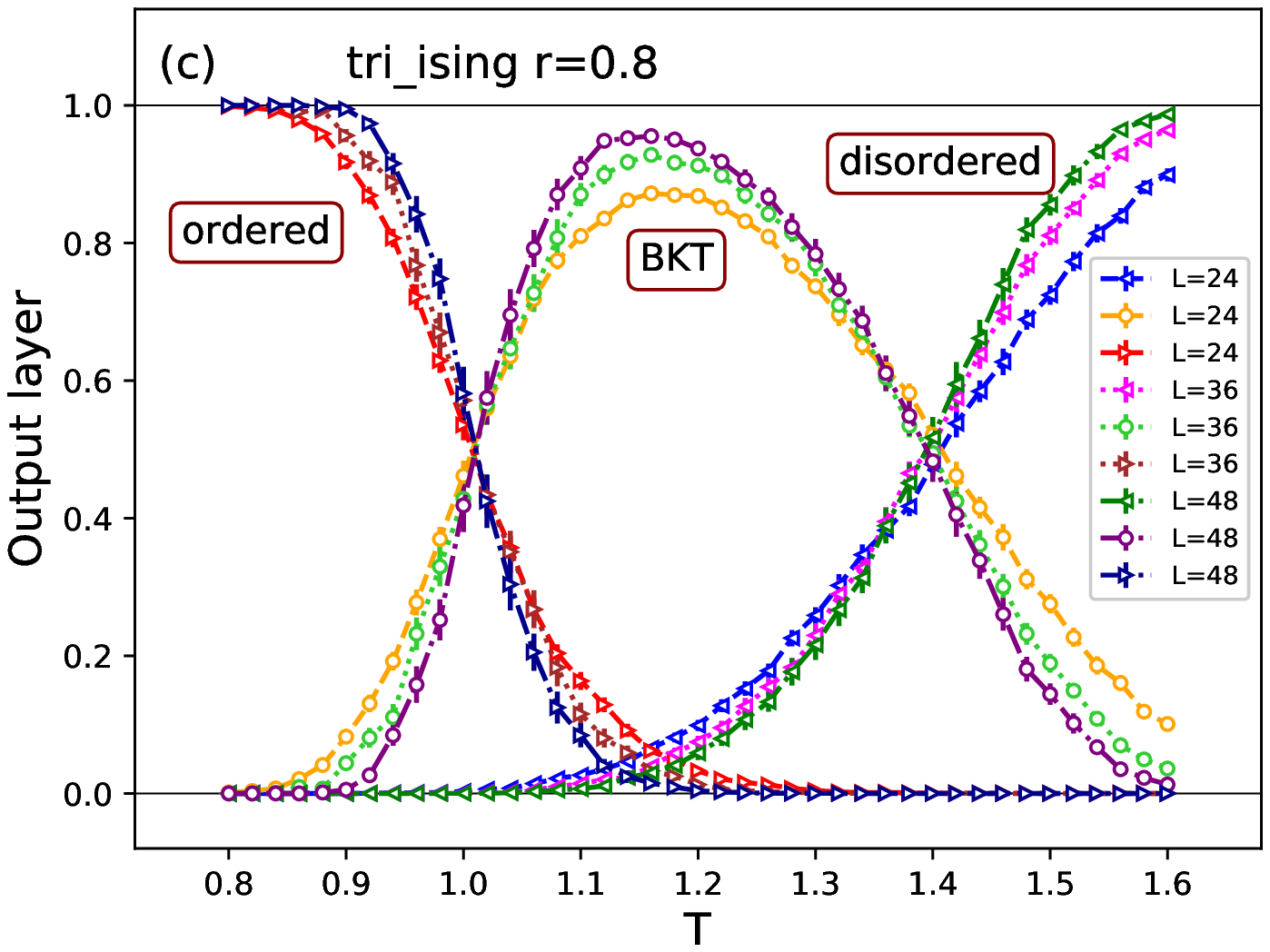}
    \caption{
      The output layer for the triangular AF Ising model
      with anisotropic NNN coupling
      for $r=0.2$ (a), $r=0.4$ (b), and $r=0.8$ (c),
      using the training data of the F six-state clock model.
      The system sizes are $L$ = 24, 36, and 48.
    }
    \label{fig:output_tri_ising}
  \end{center}
\end{figure}

As the result of the machine-learning study,
we show the output layer averaged over a test set as a function of $T$
for the triangular AF Ising model with anisotropic NNN coupling
in \myfig~\ref{fig:output_tri_ising}.
For the training data, we use the F six-state clock model on the
triangular lattice, which will be shown in the Appendix.
The data for $r = 0.2$, 0.4, and 0.8 are given in (a), (b), and (c),
respectively. The system sizes are 24, 36, and 48.
The probabilities of predicting the phases
are plotted at each temperature.
Figure~\ref{fig:output_tri_ising} clearly demonstrates the existence
of three phases, the ordered phase, the intermediate BKT phase,
and the disordered phase.
The rough estimates of the transition temperatures,
$T_1$ and $T_2$, are tabulated in \mytable~\ref{table_tri_ising}.

%%%%%%%%%%%%%%%%%%%%%%%%%%%%%%%%%%%%%%%%%%%%%%%%%%%%%%%%%%%%%%%%%%%%%%%%%%%%
%%%%%%%%%%%%%%%%%%%%%%%%%%%%%%%%%%%%%%%%%%%%%%%%%%%%%%%%%%%%%%%%%%%%%%%%%%%%

%%%%%%%%%%%%%%%%%%%%%%%%%%%%%%%%%%%%%%%%%%%%%%%%%%%%%%%%%%%%%%%%%%%%%%%%%%%%
\begin{figure}
  \begin{center}
    \includegraphics[width=5.0cm]{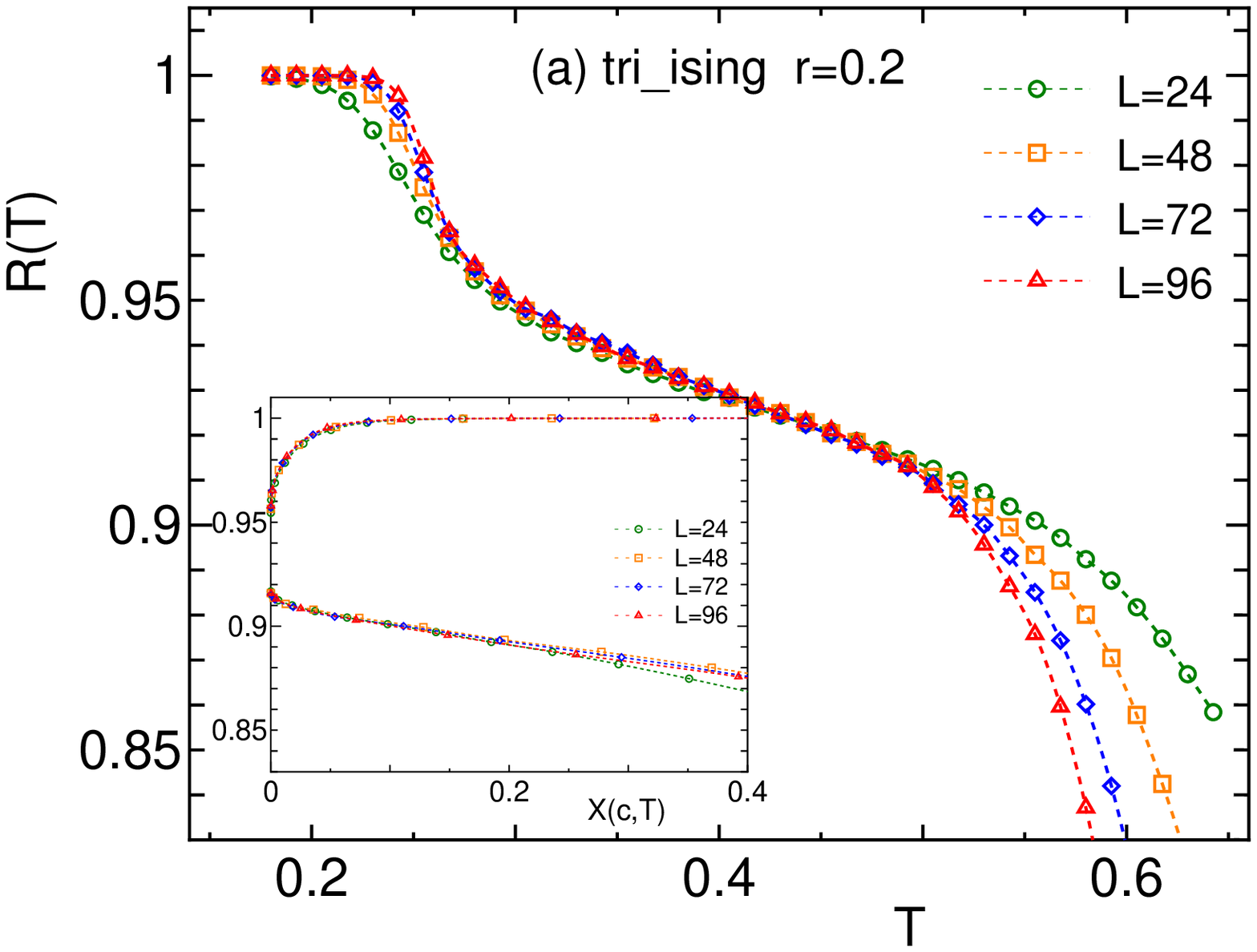}\hfill
    \includegraphics[width=5.0cm]{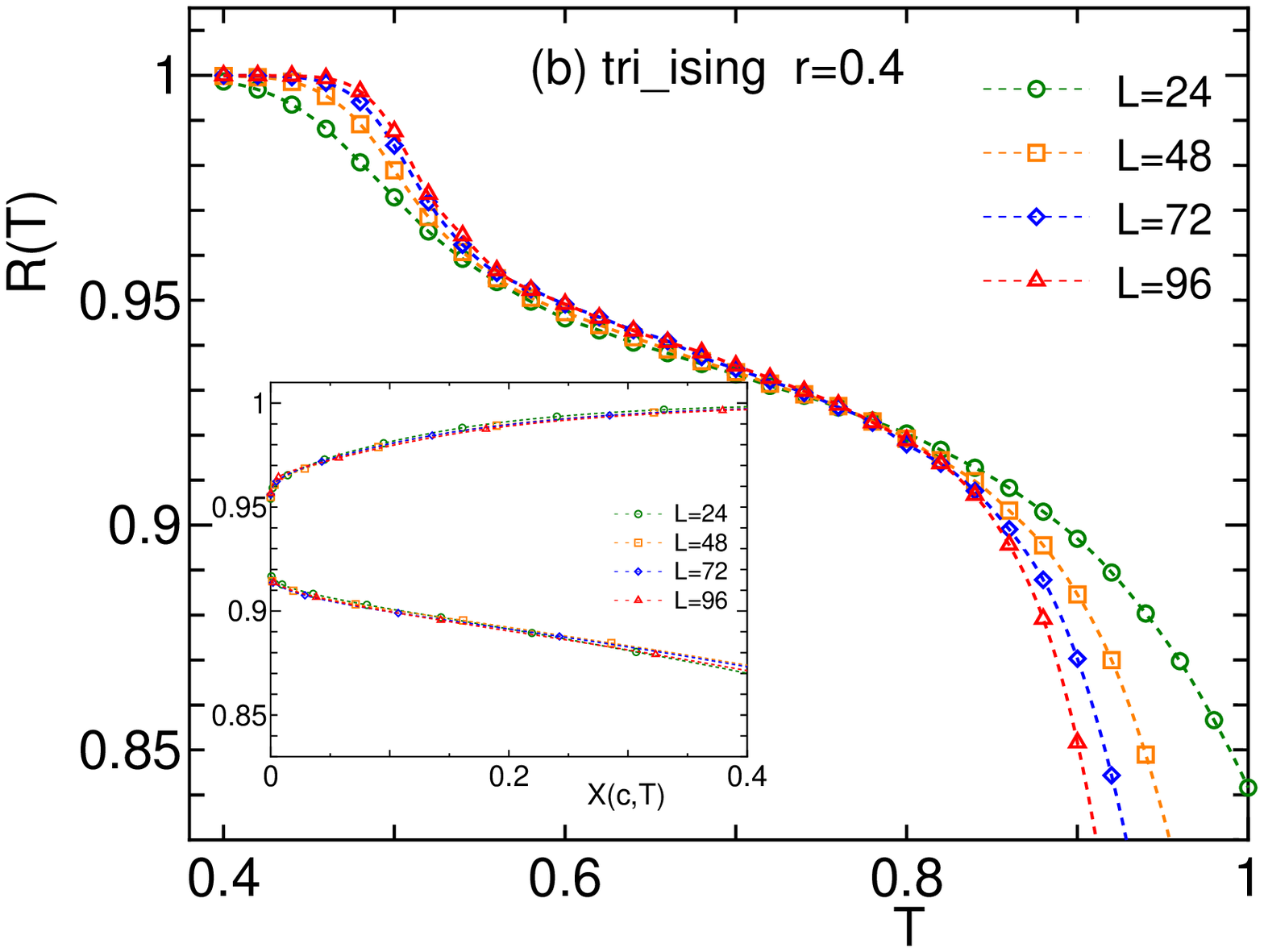}\hfill
    \includegraphics[width=5.0cm]{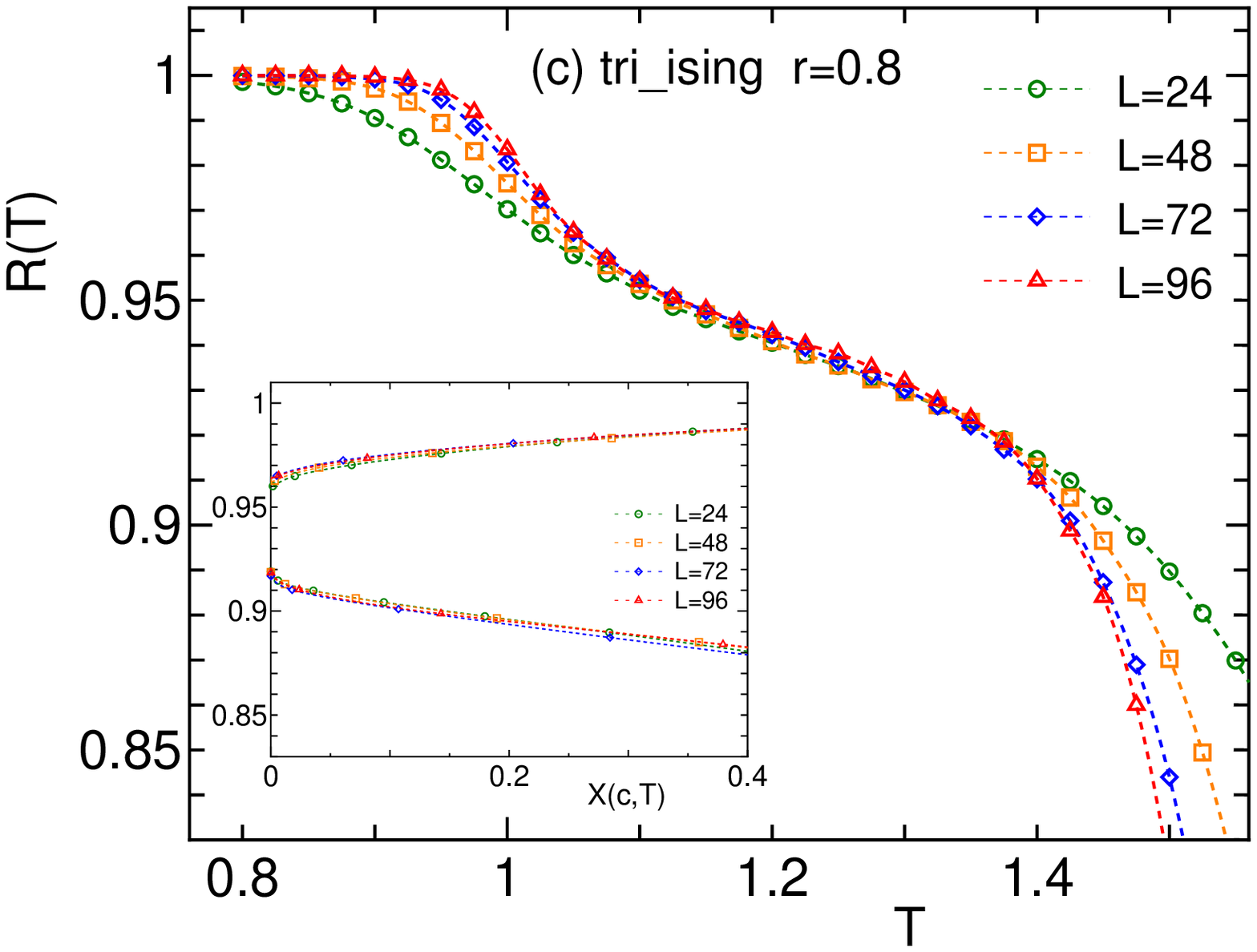}
    \caption{
      The plot of the correlation ratio $R(T)$ for the triangular
      AF Ising model with anisotropic NNN coupling.
      In the inset, the FSS plots are given,
      where $X(c,T)=L/\exp(c_{1,2}/\sqrt{|T-T_{1,2}|})$.
    }
    \label{fig:MC_tri_ising}
  \end{center}
\end{figure}
%%%%%%%%%%%%%%%%%%%%%%%%%%%%%%%%%%%%%%%%%%%%%%%%%%%%%%%%%%%%%%%%%%%%%%%%%%%%

\subsection*{Monte Carlo study}

We show the MC results of the correlation ratio,
$R(T)$, of the triangular AF Ising model
with anisotropic NNN coupling. The results for $r = 0.2$, 0.4,
and 0.8 are plotted in \myfig~\ref{fig:MC_tri_ising}.
The system sizes are $L=24$, 48, 72, and 96.
We again observe the collapsing curves of different sizes
at intermediate temperature regimes
and the spray out at lower and higher temperatures.
The behavior of the collapsing in the BKT phase is better than the case of the AF three-state Potts model
on the square lattice with F NNN coupling (\myfig~\ref{fig:MC_AFPotts}).

The FSS plots based on the exponential divergence of the correlation length
are given in the inset of \myfig~\ref{fig:MC_tri_ising},
where $T$ is scaled as $L/\exp(c_{1,2}/\sqrt{|T-T_{1,2}|})$.

The rough estimates of $T_1$ and $T_2$
are also tabulated in \mytable~\ref{table_tri_ising}.

\subsection*{Level-spectroscopy study}

\begin{figure}
  \begin{center}
    \includegraphics[width=7cm]{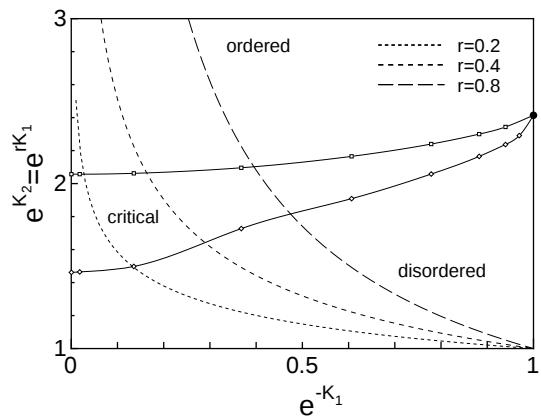}
    \caption{
    The phase diagram of the triangular AF Ising model
    with anisotropic NNN coupling
    using the level-spectroscopy method \cite{Otsuka2}
    in the space of $(u,v)$ = $(e^{-K_1}, e^{K_2})$.
    The curves $v=e^{-r\ln (u)}$ with $r$ = 0.2, 0.4, and 0.8
    are also plotted.
    From the crossing point with these curves,
    we can obtain the estimates of $T_1$ and $T_2$.
    }
    \label{fig:otsuka_tri_ising}
  \end{center}
\end{figure}

Otsuka~{\it et al.}~\cite{Otsuka2} studied this model
using the LS method.
The phase diagram, in the space of $u = e^{-K_1} = e^{-J_1/T} $ and
$v = e^{K_2} = e^{r J_1/T}$,
is reproduced in \myfig~\ref{fig:otsuka_tri_ising}, where
the curves $v=e^{-r\ln (u)}$ with $r = 0.2$, 0.4, and 0.8 are added.
From the crossing point with the curve $v=e^{-r\ln (u)}$,
we can obtain the estimates of $T_1$ and $T_2$.
The estimates are tabulated in \mytable~\ref{table_tri_ising}.

%%%%%%%%%%%%%%%%%%%%%%%%%%%%%%%%%%%%%%%%%%%%%%%%%%%%%%%%%%%%%%%%%%%%%%%%%%%%
%%%%%%%%%%%%%%%%%%%%%%%%%%%%%%%%%%%%%%%%%%%%%%%%%%%%%%%%%%%%%%%%%%%%%%%%%%%%

%%%%%%%%%%%%%%%%%%%%%%%%%%%%%%%%%%%%%%%%%%%%%%%%%%%%%%%%%%%%%%%%%%%%%%%%%%%%
\begin{table}
  \caption{
    The estimates of $T_1$ and $T_2$ for the triangular AF Ising model
    with anisotropic NNN coupling. The estimates of
    the level-spectroscopy method are based on the calculation in \cite{Otsuka2}.
  }
  \label{table_tri_ising}
  \begin{center}
    \begin{tabular}{lllllll}
      \hline
      \hline
                         & \multicolumn{2}{l}{$r=0.2$}
                         & \multicolumn{2}{l}{$r=0.4$}
                         & \multicolumn{2}{l}{$r=0.8$}                                                                       \\
                         & \ $T_1$                     & \ $T_2$                     & \ $T_1$ & \ $T_2$ & \ $T_1$ & \ $T_2$ \\
      \hline
      machine-learning   & \ 0.25                      & \ 0.48                      & \ 0.51  & \ 0.82  & \ 1.01  & \ 1.39  \\
      Monte Carlo        & \ 0.29                      & \ 0.465                     & \ 0.57  & \ 0.795 & \ 1.08  & \ 1.36  \\
      level-spectroscopy & \ 0.271                     & \ 0.494                     & \ 0.547 & \ 0.801 & \ 1.070 & \ 1.330 \\
      \hline
      \hline
    \end{tabular}
  \end{center}
\end{table}
%%%%%%%%%%%%%%%%%%%%%%%%%%%%%%%%%%%%%%%%%%%%%%%%%%%%%%%%%%%%%%%%%%%%%%%%%%%%

We see the consistency of the estimates of $T_1$ and $T_2$  by
various methods tabulated in \mytable~\ref{table_tri_ising}.
We may say that the machine-learning study of this model yields
reasonable estimates.

\section{Effective exponents}

In this section, we make a comment on the calculation of
the decay exponent $\eta$ in the MC study,
and discuss the self-dual point.

The correlation function with the distance $r$ decays as
$g(r) \sim r^{-\eta(T)}$ at the critical point (line).
We consider the correlation function with the distance $L/2$
for the finite system of the size $L$
\begin{equation}
  g_2(L) = \l \bm{s}(i) \cdot \bm{s}(i+L/2) \r.
\end{equation}
We also consider
\begin{equation}
  g_4(L) = \l {\bm s}(i) \cdot {\bm s}(i+L/4) \r.
\end{equation}
At the critical point (line),
$g_2(L)$ behaves as $L^{-\eta(T)}$
and for $T<T_1$ and $T>T_2$,
we may have the following relation:
\begin{equation}
  g_2(L) \sim L^{-\eta(T)}f(L/\xi).
\end{equation}
We examine the ratio of $g_2$ of the two sizes $L$ and $2L$:
\begin{equation}
  \frac{g_2(L)}{g_2(2L)} = \frac{L^{-\eta}f(L/\xi)}{(2L)^{-\eta}f(2L/\xi)}
  = 2^{-\eta} \frac{f(L/\xi)}{f(2L/\xi)}.
\end{equation}
Then, we define the effective exponent $\eta_{\rm eff}$ by
\begin{equation}
  \eta_{\rm eff} = \frac{\ln(g_2(L)/g_2(2L))}{\ln(2)}.
  \label{effective_exponent}
\end{equation}

\begin{figure}[t]
  \begin{center}
    \includegraphics[width=5.0cm]{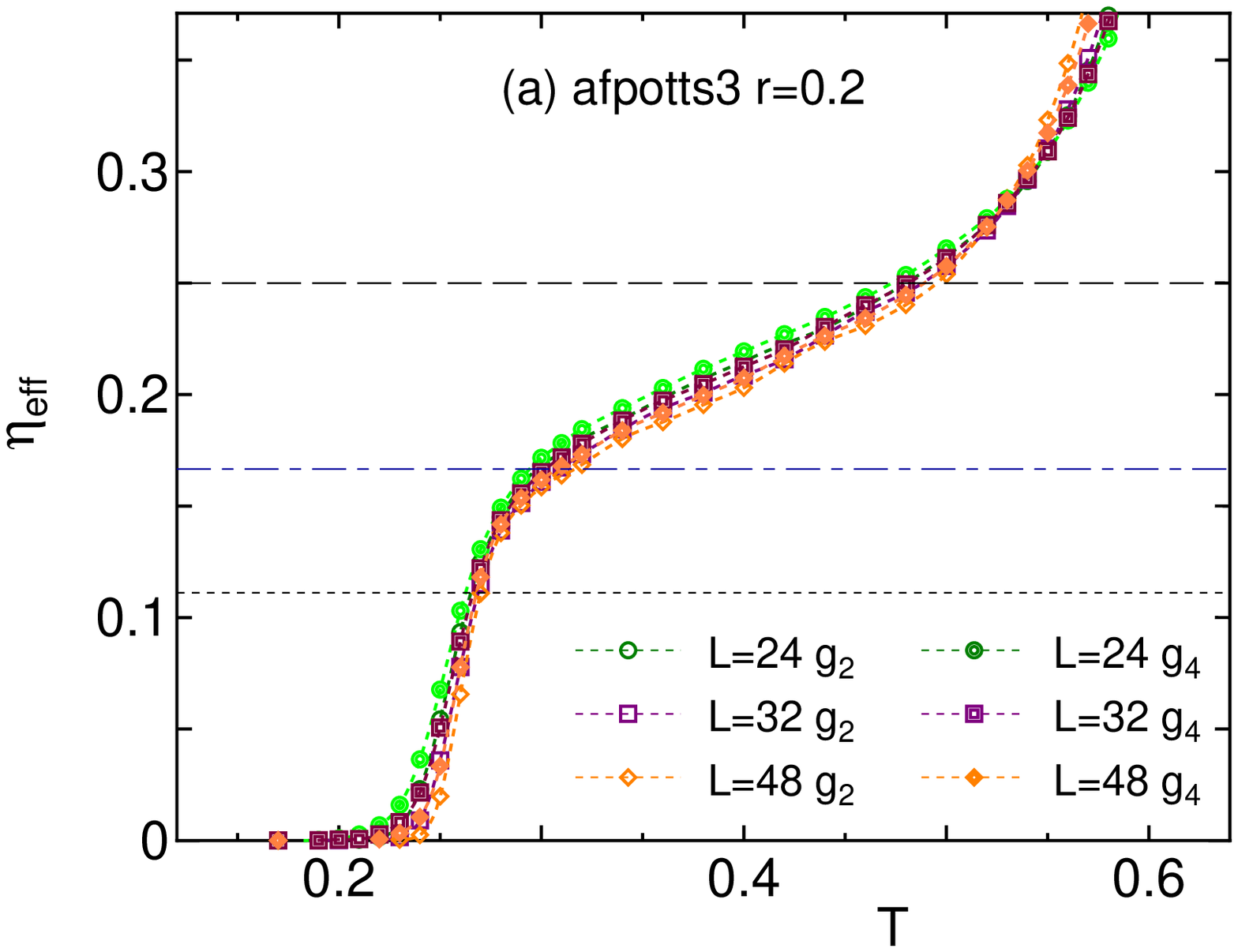}\hfill
    \includegraphics[width=5.0cm]{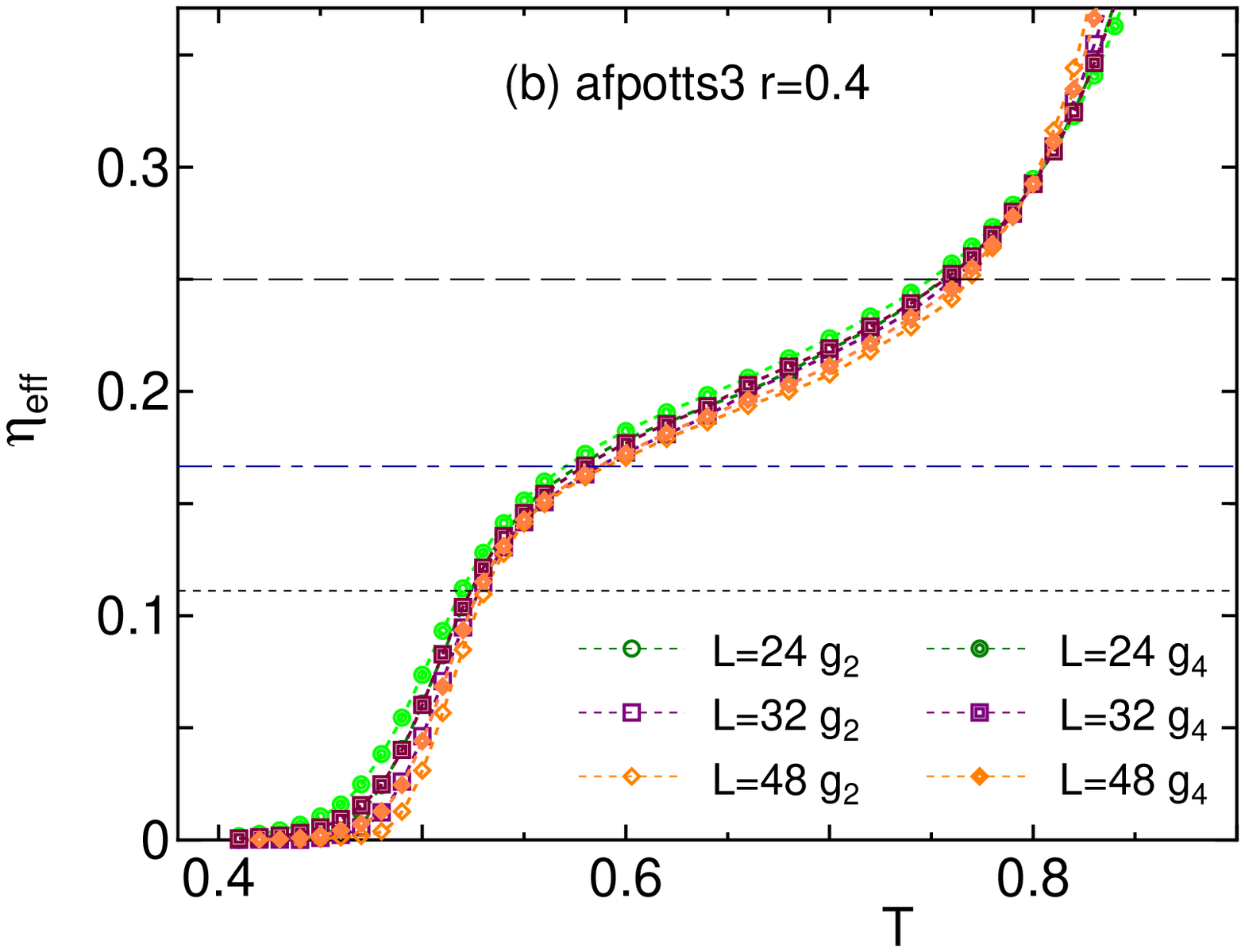}\hfill
    \includegraphics[width=5.0cm]{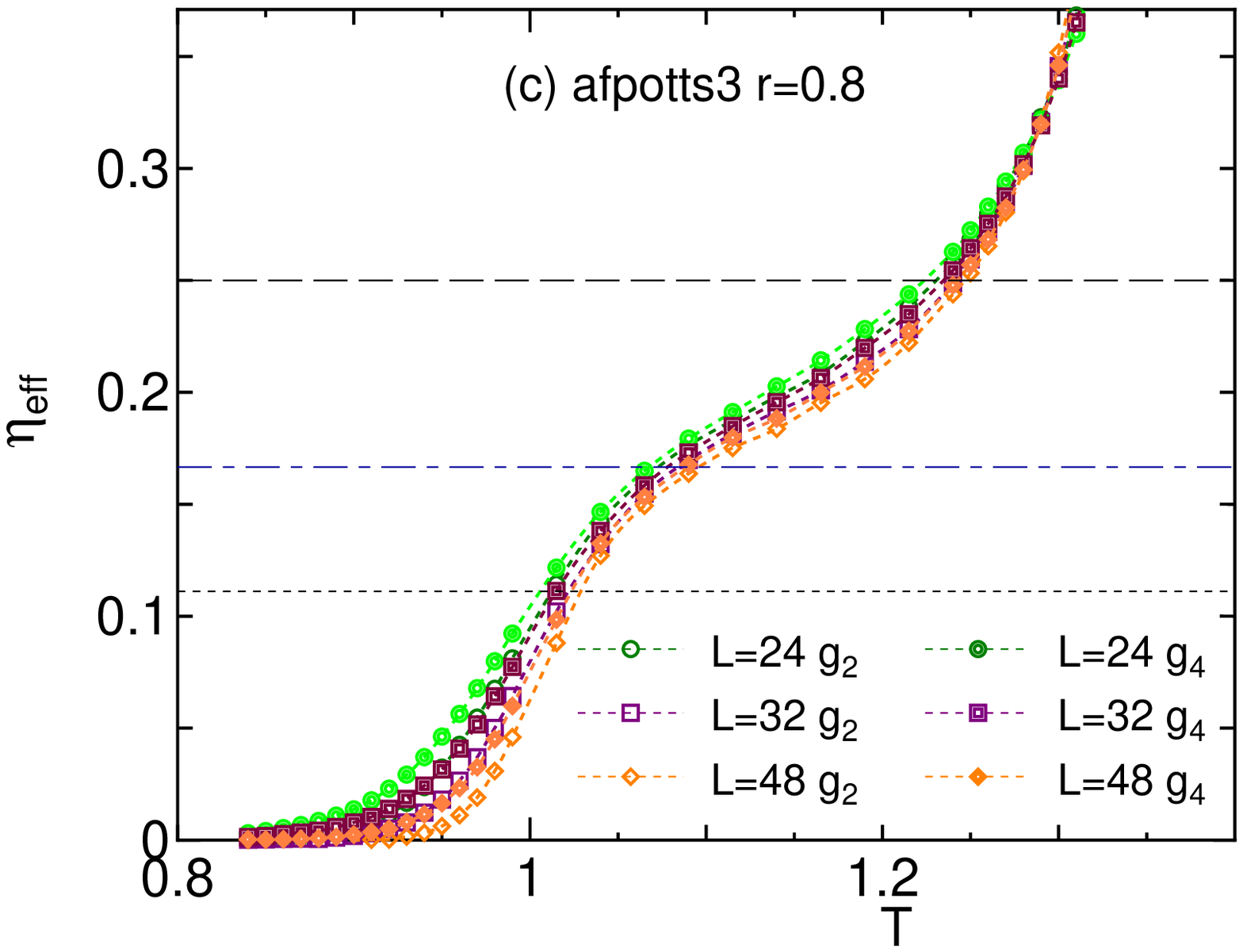}
    \caption{
      The effective exponents $\eta_{\rm eff}$, calculated
      by \myeq{\ref{effective_exponent}}, for AF three-state Potts model
      with F NNN coupling (square lattice).
      The theoretical values of $\eta_2=1/4$, $\eta_{\rm sd}=1/6$, and
      $\eta_1=1/9$, are shown by the dashed, dash-dotted, and dotted lines,
      respectively.
    }
    \label{fig:MC_AFPotts2}
  \end{center}
\end{figure}

\begin{figure}[t]
  \begin{center}
    \includegraphics[width=5.0cm]{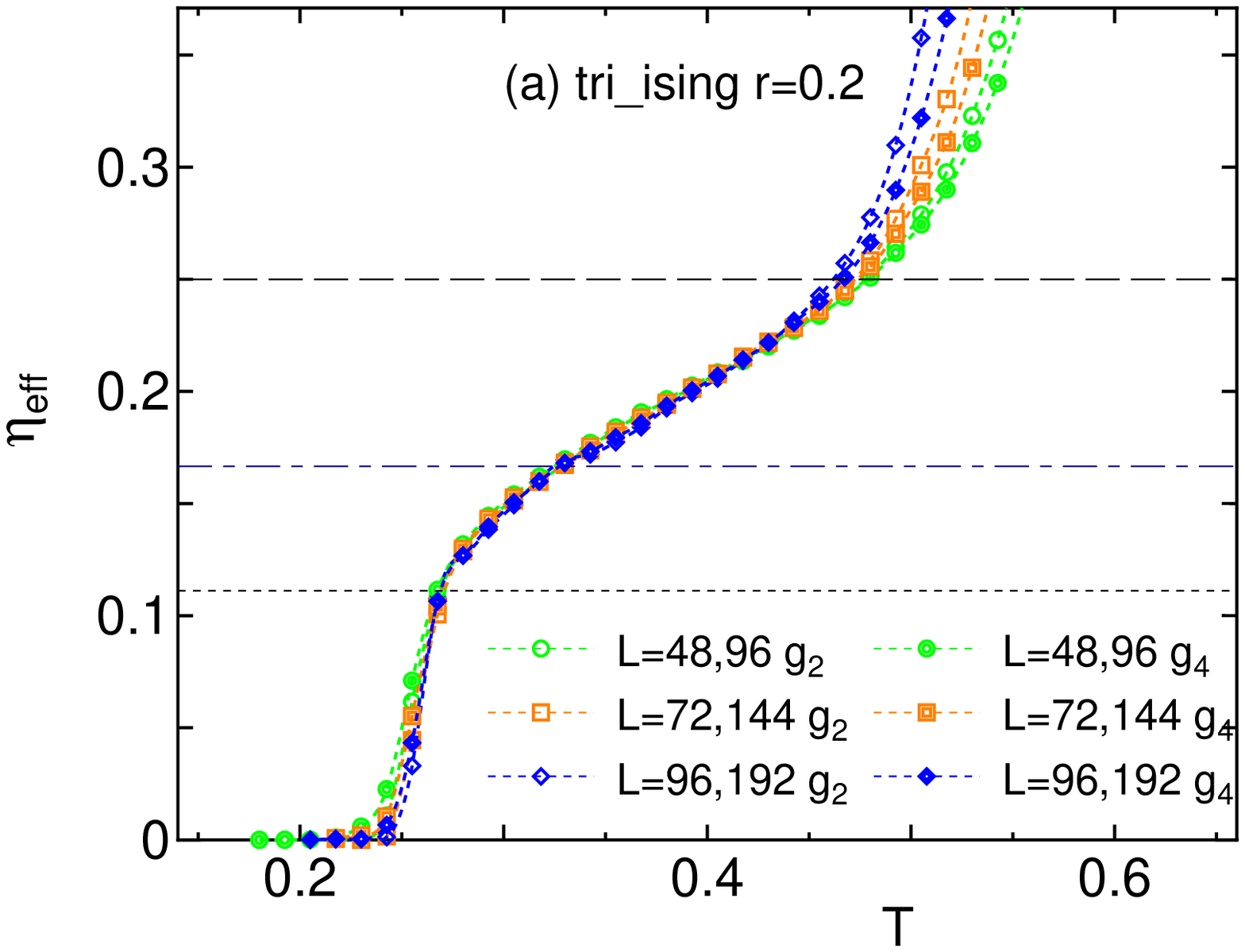}\hfill
    \includegraphics[width=5.0cm]{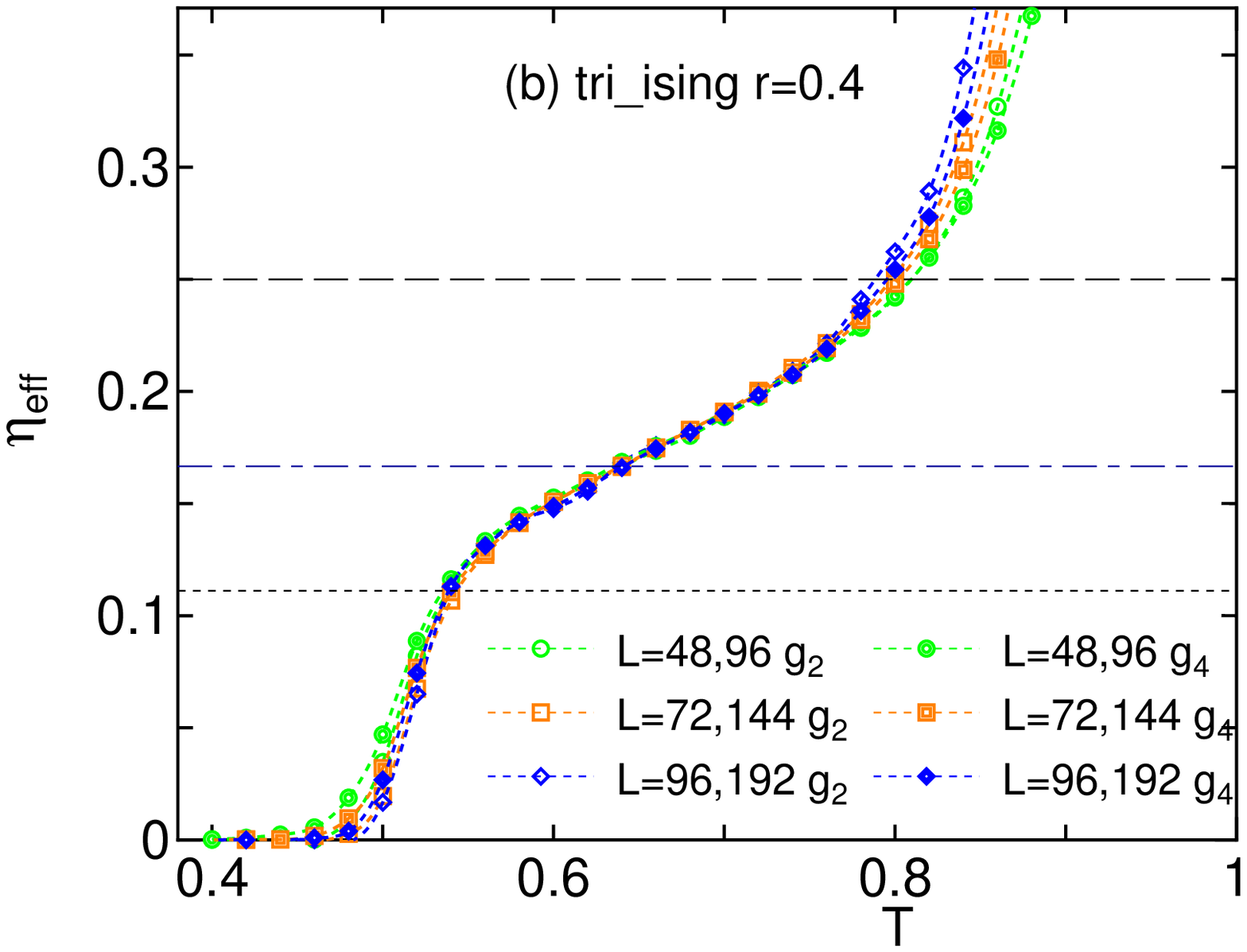}\hfill
    \includegraphics[width=5.0cm]{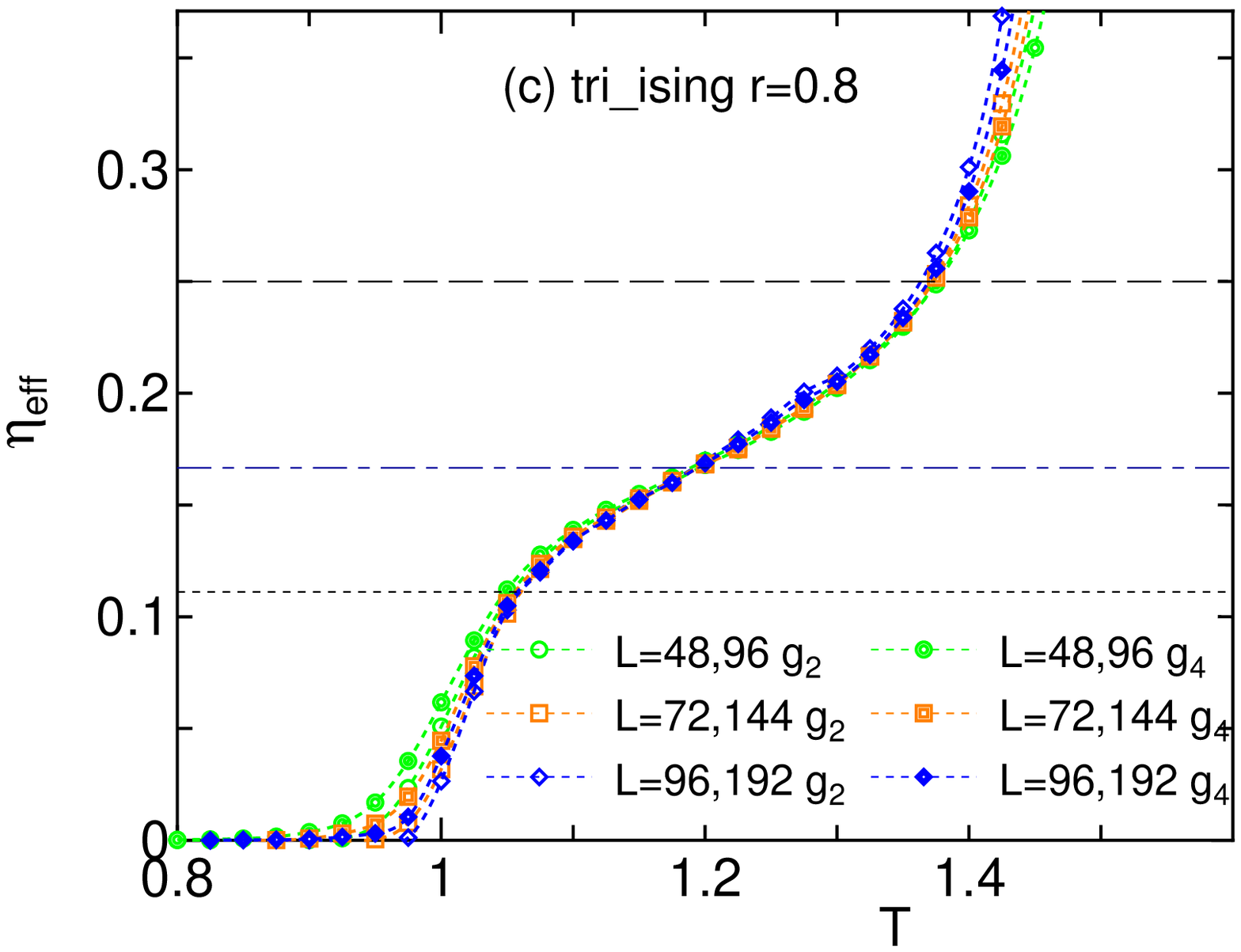}
    \caption{
      The effective exponents $\eta_{\rm eff}$, calculated
      by \myeq{\ref{effective_exponent}}, for the AF Ising model
      with anisotropic NNN coupling (triangular lattice).
      The theoretical values of $\eta_2=1/4$, $\eta_{\rm sd}=1/6$, and
      $\eta_1=1/9$ are shown by the dashed, dash-dotted, and dotted lines,
      respectively.
    }
    \label{fig:MC_tri_ising2}
  \end{center}
\end{figure}

Theoretically, the critical exponent $\eta$ becomes 1/4 at $T_2$ and
$4/q^2$ at $T_1$. We may estimate $T_2$ and $T_1$ from the condition
that $\eta_{\rm eff}$ takes the theoretical value.
The exponent $\eta_{\rm eff}$ is expected to take
the temperature-dependent value for $T_1\le T\le T_2$;
it does not depend on $L$, and there is no difference
between $g_2$ and $g_4$.  For $T<T_1$ and $T>T_2$,
it becomes $L$-dependent, and $g_2$ and $g_4$ take
different values.
We show the calculated results of $\eta_{\rm eff}$ using
\myeq{\ref{effective_exponent}} for the AF three-state Potts model
with ferromagnetic NNN coupling in \myfig~\ref{fig:MC_AFPotts2} and
for the AF Ising model with anisotropic NNN coupling on the triangular lattice
in \myfig~\ref{fig:MC_tri_ising2}, respectively.
There, the theoretical values of $\eta_2=1/4$ and
$\eta_1=1/9$ are shown by the dashed and dotted lines,
respectively.
Not only the data for $g_2$ but those for $g_4$ are given.
We see an expected behavior of the temperature dependence
of the effective exponent $\eta_{\rm eff}$ in \myfigs~\ref{fig:MC_AFPotts2}
and \ref{fig:MC_tri_ising2}.
The temperatures at which $\eta_{1,2}$ take the theoretical values
are consistent with the estimates of $T_1$ and $T_2$
in \mytables~\ref{table_AFPotts} and \ref{table_tri_ising}.

In the Villain model, where the exact duality transformation holds,
the exponent $\eta$ becomes $1/q$ at the self-dual (sd) point.
Moreover, there is an exact relation between the temperatures:
\begin{equation}
  T_1 T_2 = 4 \pi^2/q^2, \quad {\rm therefore}, \quad T_{\rm sd} = 2\pi/q.
\end{equation}
Although this self-dual point is particular to the Villain model,
we give the value of $\eta_{\rm eff}$, $1/q$, in \myfigs~\ref{fig:MC_AFPotts2}
and \ref{fig:MC_tri_ising2}, by the dash-dotted lines.
Let us denote the temperature $T_{\rm sd}$ by the condition
that $\eta_{\rm eff}$ takes the self-dual value $\eta_{\rm sd}$.
We can see that $T_{\rm sd}$ is close to the temperature
at which the probability of predicting the BKT phase takes the maximum
in Figs.~\ref{fig:output_AFpotts3} and \ref{fig:output_tri_ising}
in the machine-learning studies.
We note that for the triangular AF Ising model with anisotropic NNN coupling,
the self-dual temperature was discussed
using the LS method~\cite{Otsuka2}.

%%%%%%%%%%%%%%%%%%%%%%%%%%%%%%%%%%%%%%%%%%%%%%%%%%%%%%%%%%%%%%%%%%%%%%%%%%%%
%%%%%%%%%%%%%%%%%%%%%%%%%%%%%%%%%%%%%%%%%%%%%%%%%%%%%%%%%%%%%%%%%%%%%%%%%%%%

%%%%%%%%%%%%%%%%%%%%%%%%%%%%%%%%%%%%%%%%%%%%%%%%%%%%%%%%%%%%%%%%%%%%%%%%%%%%
%%%%%%%%%%%%%%%%%%%%%%%%%%%%%%%%%%%%%%%%%%%%%%%%%%%%%%%%%%%%%%%%%%%%%%%%%%%%

\section{Summary and discussion}

We used the machine-learning study for the AF three-state Potts model
on the square lattice with F NNN coupling
and the triangular AF Ising model with anisotropic NNN coupling.
We classified the ordered, the BKT, and the disordered states
using the training data of the F six-state clock model.
We explicitly showed the universal behavior of
implicit symmetries in totally different models.

We also used the MC studies paying attention
to the correlation ratio, which is appropriate
in studying the BKT transitions.
In the recent publication on the antiferromagnetic Potts
model~\cite{Zhang}, the authors pointed out that
the Binder ratio is not necessarily the best method
in determining the critical point as in
Refs.~\cite{katzgraber} and \cite{hasenbusch}.
The two BKT transitions were determined with precision
using the valid metric of correlation ratios, confirming
the six-state clock universality.

Comparing the results of the machine-learning study
and the MC simulation with the previous results of
the LS method~\cite{Otsuka,Otsuka2},
we confirmed the six-state clock universality
for the AF three-state Potts model on the square lattice
with F NNN coupling and the triangular AF Ising model
with anisotropic NNN coupling.

%\textcolor{red}{
We make a brief comment on a future study. Surface effects are 
important subjects of critical phenomena, and the extraordinary-log phase 
has captured recent attention.  With this regard, recently 
the three-dimensional six-state clock model~\cite{Zou} and 
the three-dimensional three-state antiferromagnetic Potts model~\cite{Zhang} 
with a free surface were studied. The application of the present approaches 
to these models will be interesting problems. 
%}

Universality is an essential concept in phase transitions.
We performed a comprehensive study on the six-state
clock universality in completely different models.
In future research, we would like to explore the six-state clock
universality in other models.

\ack
The authors thank Hiroyuki Mori and Hwee Kuan Lee
for valuable discussions.
This work was supported by JSPS KAKENHI Grant Number JP22K03472.

\section*{Appendix. $\bm{XY}$ and six-state clock models on the triangular lattice}

The reports on the {\XY} or clock models on the triangular lattice
have not been found in the literature.
Only the exception is the MC study of the {\XY} model
by Sorokin \cite{Sorokin}. By using the MC study of the
helicity modulus, the BKT temperature was estimated as 1.418(2).
We here show our MC results on the {\XY} model and the six-state
clock model on the triangular lattice.
We also perform a machine-learning study.

%%%%%%%%%%%%%%%%%%%%%%%%%%%%%%%%%%%%%%%%%%%%%%%%%%%%%%%%%%%%%%%%%%%%%%%%%%%%
\begin{figure}[t]
  \begin{center}
    \includegraphics[width=5.0cm]{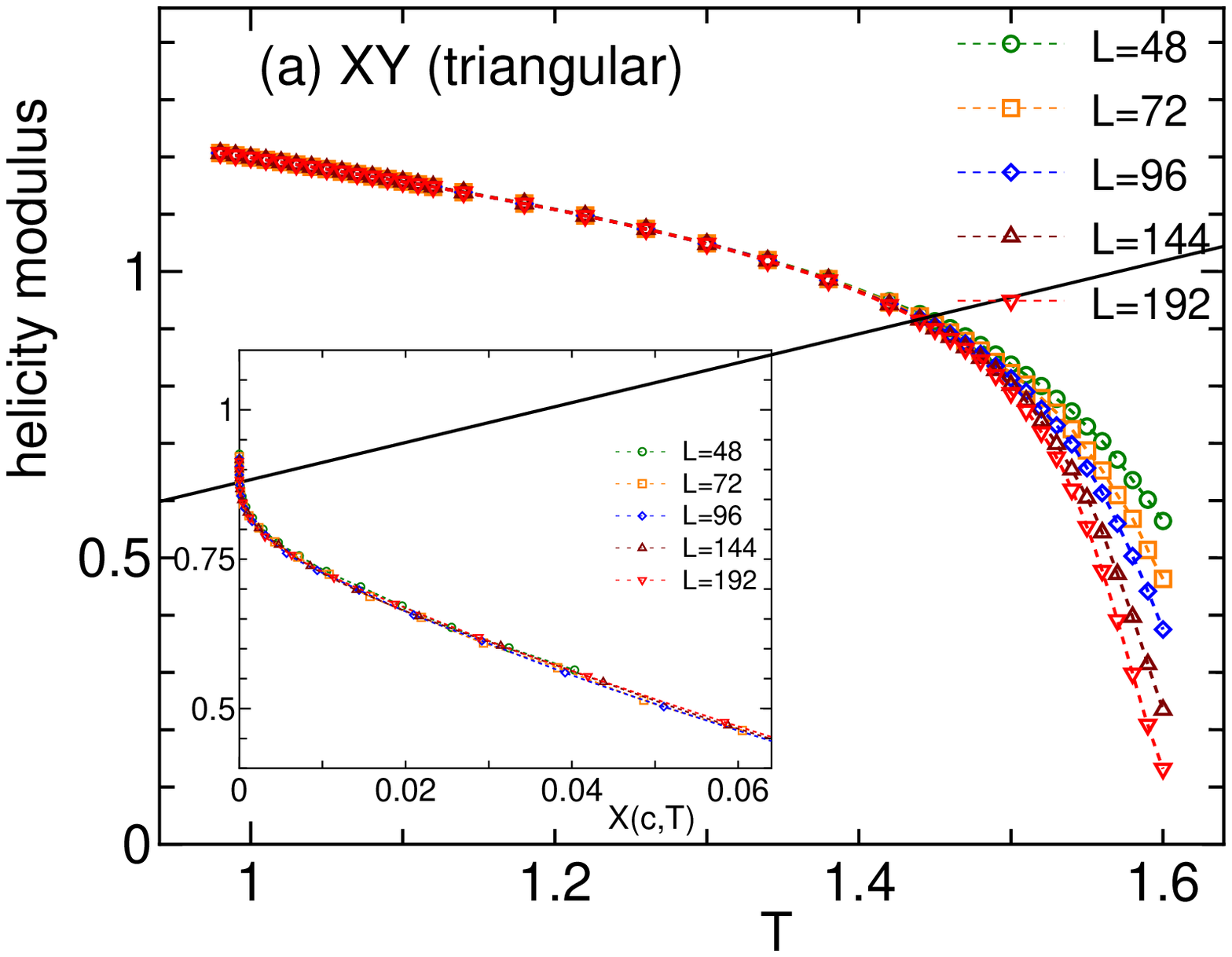}\quad
    \includegraphics[width=5.0cm]{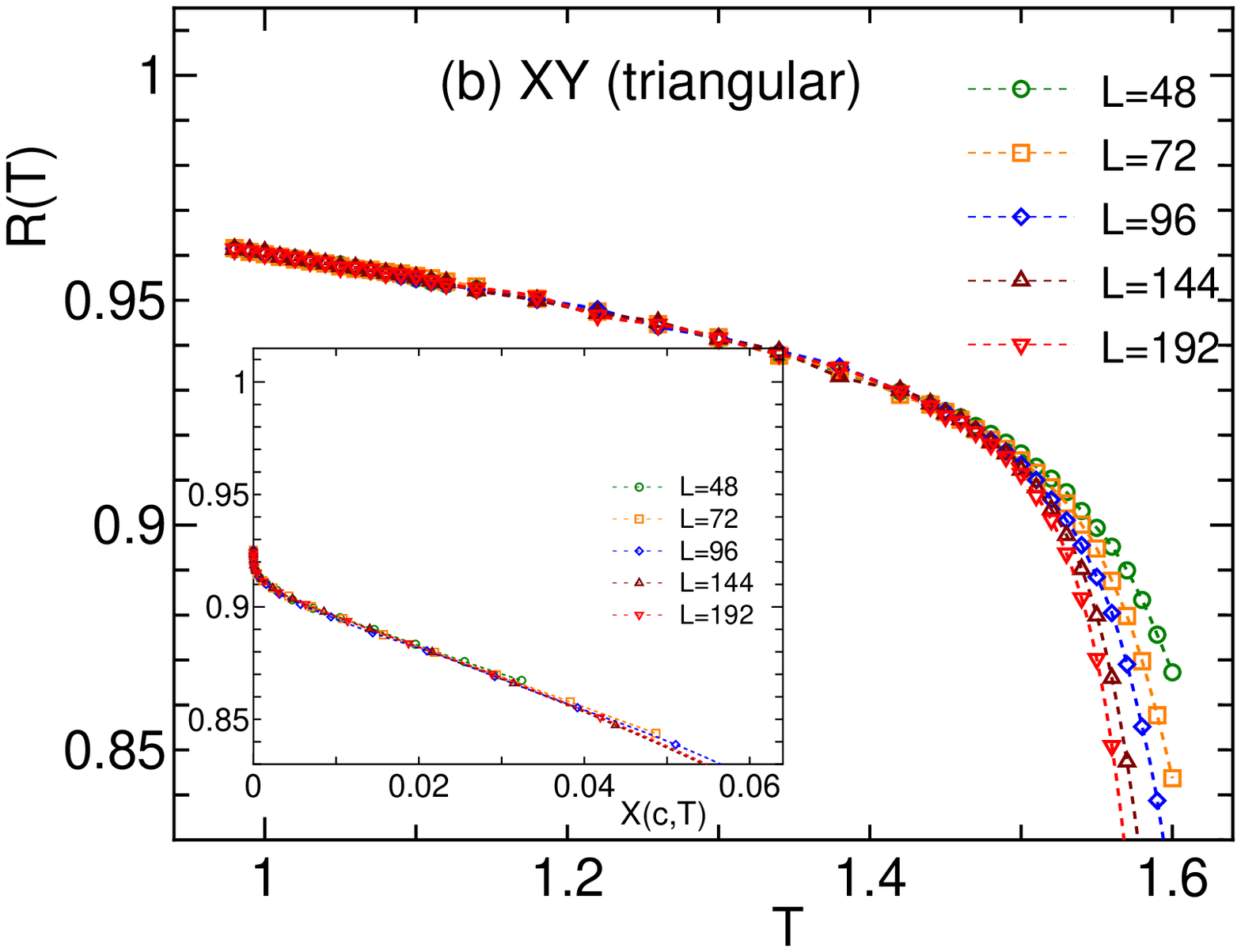}
    \caption{
      Helicity modulus (a) and correlation ratio (b) for {\XY} model
      on the triangular lattice.
      The system sizes are 48, 72, 96, 144, and 192.
      In the plot of helicity modulus, we give the straight line $(2/\pi) T$.
      In the inset the FSS plots are given,
      where $X(c,T)=L/\exp(c_{2}/\sqrt{|T-T_{2}|})$.
    }
    \label{fig:MC_tri_XY}
  \end{center}
\end{figure}
%%%%%%%%%%%%%%%%%%%%%%%%%%%%%%%%%%%%%%%%%%%%%%%%%%%%%%%%%%%%%%%%%%%%%%%%%%%%

\begin{figure}[t]
  \begin{center}
    \includegraphics[width=5.0cm]{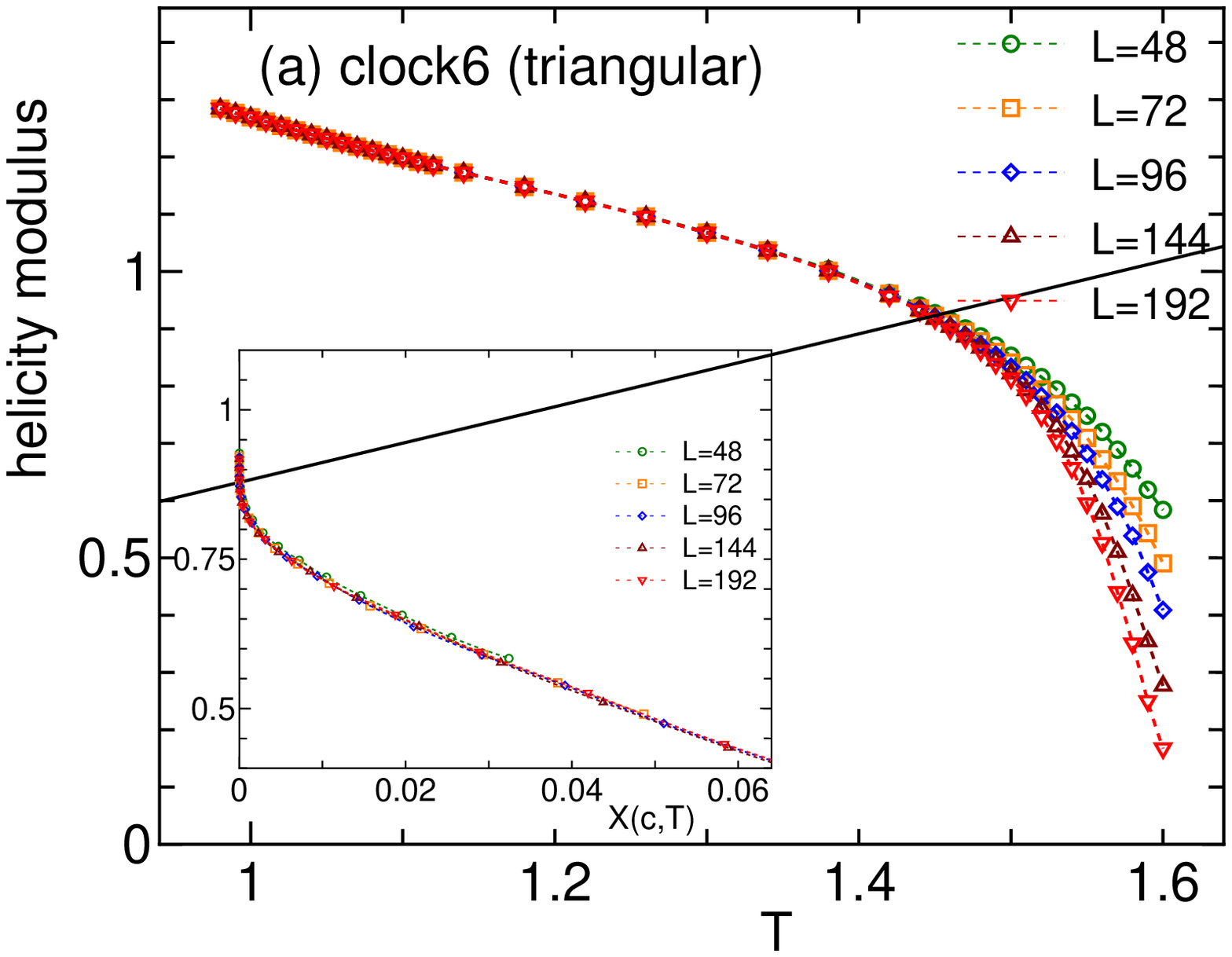}\quad
    \includegraphics[width=5.0cm]{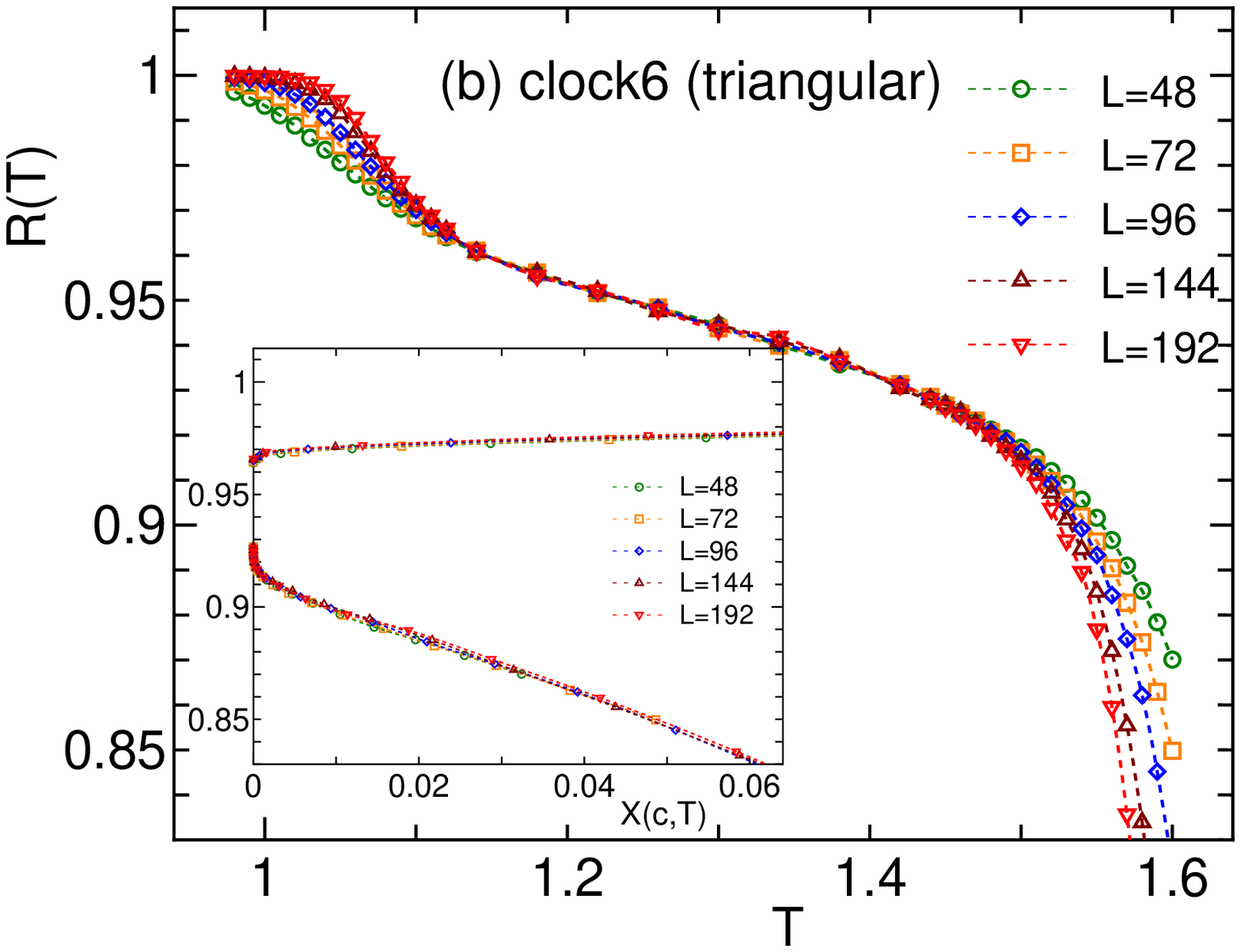}
    \caption{
      Helicity modulus (a) and correlation ratio (b) for the six-state clock model
      on the triangular lattice.
      The system sizes are 48, 72, 96, 144, and 192.
      In the plot of helicity modulus, we give the straight line $(2/\pi) T$.
      In the inset the FSS plots are given,
      where $X(c,T)=L/\exp(c_{1,2}/\sqrt{|T-T_{1,2}|})$.
      For the helicity modulus, only the FSS for the high-temperature side
      is given.
    }
    \label{fig:MC_tri_clock}
  \end{center}
\end{figure}

The MC results on the {\XY} model and the six-state
clock model are plotted in Figs.~\ref{fig:MC_tri_XY} and
\ref{fig:MC_tri_clock}, respectively.
The helicity modulus and correlation ratio $R(T)$
are plotted in both figures.
The system sizes are 48, 72, 96, 144, and 192.
In the plot of helicity modulus,
we give the straight line $(2/\pi) T$. The crossing point gives
a universal jump. The $q$-state clock model, a discrete version
of the {\XY} model, experiences two BKT transitions because of
the discreteness.  We note that there is no anomaly in helicity
modulus for the lower transition, $T_1$.
The BKT phase is indicated by the collapsing curves of different sizes
in the plot of $R(T)$.  It is a direct consequence
of the power law behavior of the correlation function at the BKT phase.
The spray out of the curves at lower and higher temperatures
signifies BKT transitions.
The rough estimates of the BKT temperatures are
$T_2$ = 1.43 for the {\XY} model, and $T_2$=1.44 and $T_1$ = 1.12
for the six-state clock model.
The result of the {\XY} model is consistent with the report by
Sorokin~\cite{Sorokin}.  The values of the triangular lattice
are roughly 3/2 of those of the square lattice.
It is because the coordination number of the triangular lattice is six,
whereas that of the square lattice is four.

%%%%%%%%%%%%%%%%%%%%%%%%%%%%%%%%%%%%%%%%%%%%%%%%%%%%%%%%%%%%%%%%%%%%%%%%%%%%
%%%%%%%%%%%%%%%%%%%%%%%%%%%%%%%%%%%%%%%%%%%%%%%%%%%%%%%%%%%%%%%%%%%%%%%%%%%%

%%%%%%%%%%%%%%%%%%%%%%%%%%%%%%%%%%%%%%%%%%%%%%%%%%%%%%%%%%%%%%%%%%%%%%%%%%%%
\begin{figure}
  \begin{center}
    \includegraphics[width=5.0cm]{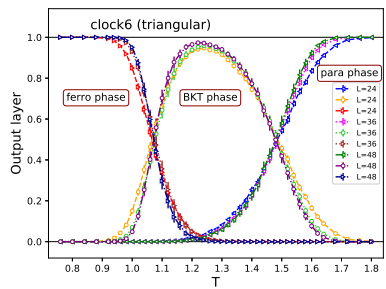}
    \caption{
      The output layer averaged over a test set as a function of $T$
      for the triangular lattice six-state clock model.
      The system sizes are $L = 24$, 36, and 48.
      The samples of $T$ within the ranges $0.76 \le T \le 1.04$,
      $1.21 \le T \le 1.31$ and
      $1.52 \le T \le 1.8$ are used for training data.
    }
    \label{fig:output_tri_clock6}
  \end{center}
\end{figure}
%%%%%%%%%%%%%%%%%%%%%%%%%%%%%%%%%%%%%%%%%%%%%%%%%%%%%%%%%%%%%%%%%%%%%%%%%%%%

We also show the results of the machine-learning study.
The output layer averaged over a test set as a function of $T$
for the triangular F six-state model is shown
in \myfig~\ref{fig:output_tri_clock6}.
The system sizes are 24, 36, and 48.
The samples of $T$ within the ranges $0.76 \le T \le 1.04$,
$1.21 \le T \le 1.31$ and
$1.52 \le T \le 1.8$ are used for the low-temperature,
middle-temperature, and high-temperature training data, respectively.
The probabilities of predicting the phases
are plotted at each temperature.
We clearly observe three phases, the ordered, the BKT, and
the disordered phases.
We estimate the size-dependent $T_{1,2}(L)$ from the point
that the probabilities of predicting two phases are 50\%.
Systematic size dependence is not observed.
The rough estimates of $T_1$ and $T_2$ are 1.07 and 1.48, respectively.
The estimates are compatible to those of the MC study.
However, the BKT phase is slightly wider, which is the same
situation as the case of the square lattice.  It comes from
the finite-size effects.

\newpage
\newcommand{\mybibitem}[6]{\bibitem{#1} #2 #6 #3 {\it #4} #5.}

\end{document}